\documentclass[aps,prl,shownopacs,twocolumn,superscriptaddress,longbibliography]{revtex4-1}
\usepackage{bm,color,amsmath,amssymb,mathrsfs,latexsym,graphicx,psfrag}

\newcommand{\braket}[2]{\left\langle #1 | #2 \right\rangle}
\newcommand{\bra}[1]{\left\langle#1\right|}
\newcommand{\ket}[1]{\left|#1\right\rangle}

\newcommand{\beq}{\begin{equation}}
\newcommand{\eneq}{\end{equation}}

\begin{document}
\title{Fractional Chern Insulator}
\author{N. Regnault}
\affiliation{Laboratoire Pierre Aigrain, ENS and CNRS, 24 rue Lhomond, 75005 Paris, France}
\author{B. Andrei Bernevig} 
\affiliation{Department of Physics, Princeton University, Princeton,
  NJ 08544}


\begin{abstract}
Chern insulators are band insulators exhibiting a nonzero Hall conductance but preserving the lattice translational symmetry. We conclusively show that a partially filled Chern insulator at $1/3$ filling exhibits a fractional quantum Hall effect and rule out charge-density wave states that have not been ruled out by previous studies. By diagonalizing the Hubbard interaction in the flat-band limit of these insulators, we show the following: The system is incompressible and has a $3$-fold degenerate ground state whose momenta can be computed by postulating an generalized Pauli principle with no more than $1$ particle in $3$ consecutive orbitals. The ground state density is constant, and equal to $1/3$ in momentum space. Excitations of the system are fractional statistics particles whose total counting matches that of quasiholes in the Laughlin state based on the same generalized Pauli principle. The entanglement spectrum of the state has a clear entanglement gap which seems to remain finite in the thermodynamic limit.  The levels below the gap exhibit counting identical to that of Laughlin $1/3$ quasiholes. Both the $3$ ground states and excited states exhibit spectral flow upon flux insertion. All the properties above disappear in the trivial state of the insulator - both the many-body energy gap and the entanglement gap close at the phase transition when the single-particle Hamiltonian goes from topologically nontrivial to topologically trivial. These facts clearly show that fractional many-body states are possible in topological insulators. 
 \end{abstract}
\date{\today}


\maketitle

\section{Introduction}

The interest in the field of topological insulators has exploded in recent years, fueled by the theoretical prediction \cite{kane-PhysRevLett.95.226801,Bernevig15122006} and experimental observation \cite{Koenig02112007,hsieh-nature2008452} of topological insulators that preserve the time reversal symmetry. These insulators have an insulating bulk but exhibit perfectly metallic edge or surface states whose existence is required by the nontrivial topology of the bulk electronic structure. However, most of the theoretical work on topological insulators is essentially of single-particle nature and neglects interactions, except for perhaps at random phase approximation level. The few interacting states \cite{pesin-naturephy2010} that have been proposed to give rise to topological insulators can be understood from a mean-field perspective \cite{rachel-PhysRevB.82.075106} - their quasiparticle excitations are still electronlike and not anyonlike. As such, the type of interacting topologically ordered state with fractional statistic quasiparticles such as the fractional quantum Hall (FQH) states, has not been found or proposed in an interacting topological insulator. The discovery of topologically ordered states of matter in the absence of the usual external magnetic field applied to the system would be of tremendous interest to the condensed matter community and could potentially revolutionize fields such as topological quantum computation.

Interestingly, it has recently been suggested \cite{sheng-natcommun.2.389,neupert-PhysRevLett.106.236804,wang-2011arXiv1103.1686W,tang-PhysRevLett.106.236802} that a Chern insulator, i.e., a zero magnetic equivalent of the quantum Hall effect \cite{haldane-1988PhRvL..61.2015H}, with strong interactions (compared to the bandwidth) has as a ground state, a fractional Hall effect at filling $1/3$ as it ground state. By diagonalizing the Hubbard interaction in a model with an almost-flat band dispersion, the authors of \cite{sheng-natcommun.2.389,neupert-PhysRevLett.106.236804,wang-2011arXiv1103.1686W} have found, for small sizes of the system, an almost $3$-fold degenerate ground state, reminiscent of the fractional quantum Hall state on a torus.  However, the origins of such a degenerate state can be either a charge-density wave or a fractional quantum Hall state. Arguments were given for why the state should be a FQH state, but they do not differentiate qualitatively between the FQH state and a charge density wave, as will be shown later. The very small sizes of the systems considered in the previous works and the lack of qualitatively unequivocal calculations that prove the existence of a fractional Chern insulator state make such a state still elusive. In the present paper, we present the proof of principle that a $\nu=1/3$ FQH state exists in the nontrivial  Chern insulator subject to a Hubbard interaction.

 By working in the flat-band limit where the kinetic energy is zero  and the band gap can be set to infinity, we present multiple results that strongly suggest the presence of a $\nu=1/3$ FQH state. We show that a repulsive interaction Hamiltonian has a $3$ fold-degenerate  ground state at filling $1/3$ separated from the excited states  by a gap which we show approaches a finite value in the thermodynamic limit. Upon the adiabatic insertion of a magnetic flux quantum, the degenerate ground states flow into each other - and cross without level repulsion at large particle number. The flux insertion and the 3-fold ground state degeneracy have already been found in two earlier papers, but, without proving the existence of a finite gap in the thermodynamic limit, they could very well be the hallmark of a charge-density wave (CDW). By going to  larger sizes than previously achieved, we show that the energy gap between the 3-fold degenerate ground state and the first excited states seems to remain finite in the thermodynamic limit  when the single-particle Hamiltonian is a nontrivial Chern insulator. Indeed, a commensurate CDW at fractional filling would also give rise to a gaped phase. Thus, the gap argument might be insufficient to prove the existence of a Laughlin-like phase. The ground states occur at lattice momenta consistent with an emergent generalized Pauli principle forbidding the presence of more than $1$ particle in $3$ consecutive orbitals. We compute the momentum space density $n(\vec{k})$ of each of the $3$ degenerate ground states and find it to be a constant approaching $1/3$ for all $\vec{k}$, another clue that the ground state is not a CDW but a true FQH state.  We then show that the quasihole excitations of these states resemble Laughlin-FQH excitations: Their total counting matches exactly that of the Laughlin quasihole states on the torus. We also present a \emph{heuristic} Pauli principle which, in some cases, counts the number of quasihole states for each momentum sector. The matching of the excitation properties is a clear signal that the state observed is a FQH state: Charge density wave states would not have an excitation spectrum resembling that of a FQH state. We then compute the entanglement spectrum (ES) for the ground state of the repulsive Hamiltonian, find a large entanglement gap, and show that the counting of the entanglement states below the gap matches that of Laughlin quasiholes. The entanglement gap remains finite in the thermodynamic limit, and closely tracks the gap in the energy spectrum as the system is tuned through a phase transition of the single-particle Hamiltonian between the nontrivial Chern insulators and a trivial band insulator.  We also discuss the dependence of the state  on the symmetries and aspect ratio of the lattice and on the parameters of the Chern insulators. Both the energy spectrum and the entanglement spectrum of the interacting spectrum change fundamentally as the single-particle Chern insulator Hamiltonian undergoes  a phase transition and becomes a trivial band insulator. In the end, we discuss the analytical principle behind the counting of quasihole states on the lattice. 

\section{The model and its symmetries}

\begin{figure}[tbp]
\includegraphics[width=3.5in]{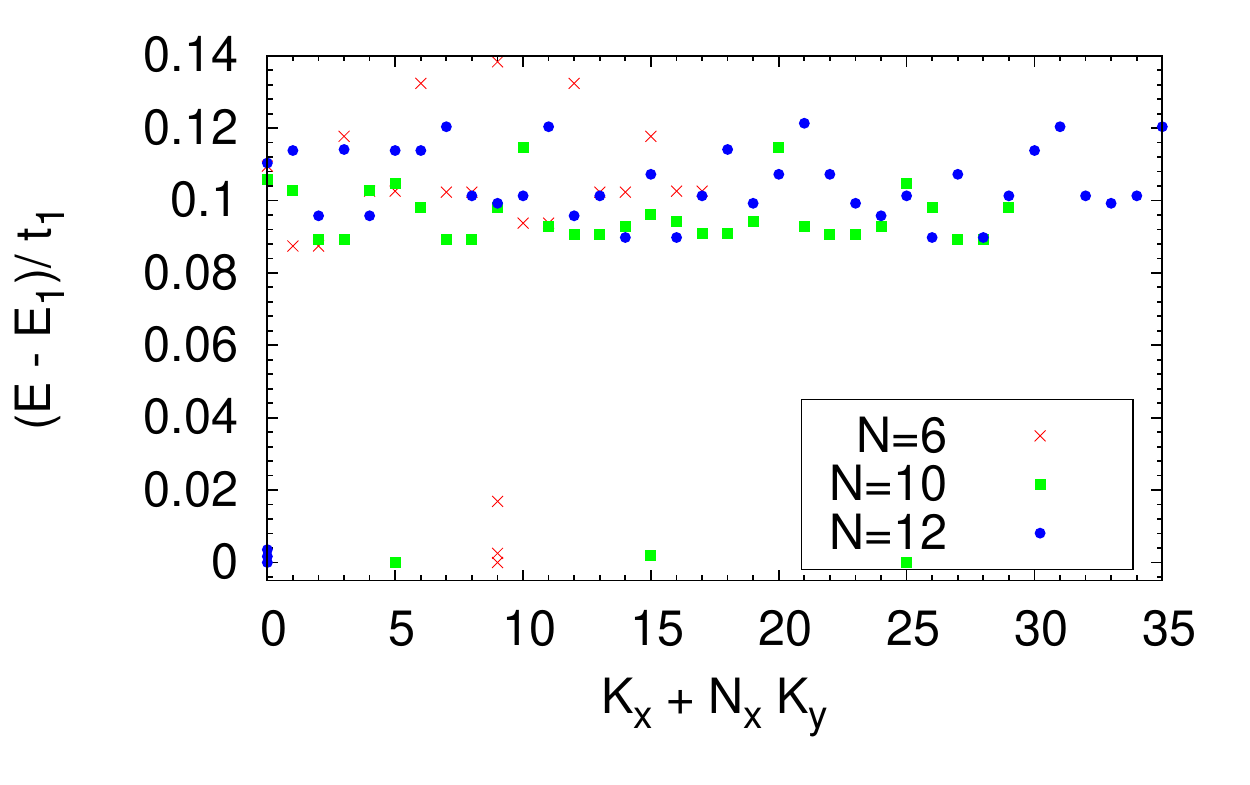}
\caption{Low energy spectrum for $N=6, 10$ and 12, $N_x=N/2, N_y=6$. The energies are shifted by $E_1$, the lowest energy for each system size. We only show the lowest energy per momentum sectors in addition to the $3$-fold ground state. We note the good ground state degeneracy, even for relatively small system sizes such as $6$ particles}\label{groundenergy}
\end{figure}

Several Chern insulator models with an approximate flat band have been recently proposed \cite{neupert-PhysRevLett.106.236804,sun-2010arXiv1012.5864S,tang-PhysRevLett.106.236802}. For our model, we pick the Chern insulator on a checkerboard lattice, first introduced in \cite{sun-2010arXiv1012.5864S,neupert-PhysRevLett.106.236804}. This model already exhibits weak dispersion of the bands, but because we work in the flat-band limit, this is not essential to our calculation - we could have picked an alternate single-particle Hamiltonian.  As written in \cite{sun-2010arXiv1012.5864S} the one-body Hamiltonian reads $H_1 = \sum_{k} (c_{kA}^\dagger , c_{kB}^\dagger) h_1(k)  (c_{kA} , c_{kB})^T$ with $A,B$ being the two sites in the unit cell. The Hamiltonian matrix can be expressed in terms of the $3$ Pauli matrices $h_1(k) = \sum_i d_i (k) \sigma_i $ where the $d_i(k)$'s are $d_x(k) = 4 t_1 \cos(\phi) \cos(k_x/2) \cos(k_y/2)$, $d_y(k) = 4 t_1 \sin(\phi) \sin(k_x/2) \sin(k_y/2)$, $d_z = 2 t_2(\cos(k_x) - \cos(k_y) ) + M$ In the original model \cite{sun-2010arXiv1012.5864S} there is an additional diagonal term -$4 t_3 \cos (k_x) cos (k_y) $ - which is of course relevant for the energy (and shrinks the dispersion of the bands, thereby making them flatter) but does not matter for the eigenstates. Since we are diagonalizing in the flat-band limit, we neglect this term.  $\phi$ is the phase factor added to the nearest neighbor hoppings, while the parameter $M$ is a mass added in order to drive the transition from a topological Chern insulator (for $M=0$) to a trivial atomic limit insulator when $M\rightarrow  \pm \infty$. The model always has an energy gap (for $t_1, t_2, \phi$ not vanishing) with the exception of the points $k_x=0, k_y=\pi, M=-4 t_2$ and $k_x=\pi, k_y =0, M=4 t_2$ which are gapless and where the phase transitions between the atomic limits $M\rightarrow  \pm \infty$ and the Chern insulator phase occur. For $|M|< 4 |t_2|$ the filled valence band has a Hall conductance of $1$. The single  particle Hamiltonian matrix has the following symmetries: inversion with identity inversion matrix $h_1(-k_x, -k_y)= h_1(k_x, k_y)$, as well as (at $M=0$) a certain type-of particle hole symmetry coupled with a $C_4$ rotation and a mirror operation:  $\sigma_z h_1(k_x, k_y) \sigma_z = - h_1(k_y, k_x)$. Unfortunately, due to the presence of fractions $k/2$, the model in \cite{sun-2010arXiv1012.5864S} is not in Bloch form. To render it in Bloch form, we perform the gauge transformation $c_{kB} \rightarrow c_{kB} \exp(-i (k_x- k_y)/2)$ to obtain:
\begin{eqnarray}
& h_2(k)= \left( {\begin{array}{cc}
 h_{11}(k) & h_{12}(k)  \\
 h_{12}^\star(k) &  - h_{11}(k) \\
 \end{array} } \right) \nonumber \\ & h_{12}(k) =  t_1e^{i \phi}(1+ e^{i(k_y-k_x)}) + t_1 e^{- i \phi}(e^{i k_y}+ e^{-i k_x}) \nonumber \\ & h_{11}(k) = 2 t_2 (\cos(k_x) - \cos(k_y)) + M
 \end{eqnarray} The inversion symmetry of $h_1(k)$ translates into another symmetry of $h_2(k)$ given by $U^\dagger(k) h_2(\vec{k}) U(k) = h_2(-\vec{k})$ with $U(k)$ a diagonal $2 \times 2$ unitary matrix with $1, e^{-i (k_x- k_y)/2}$ on the diagonal. 
 We now fractionally fill the valence band of this insulator and add interactions. The existence of a FQH state in a Chern insulator should not be taken for granted. Unlike the Landau level, a band insulator has nonzero bandwidth and a Berry phase distribution which cannot be made uniform over the full Brillouin zone. Fractionally filling the band will, even in the presence of large interactions, allow the particles to cover only part of the Brillouin zone and hence feel only part of the Berry curvature. To eliminate the effect of the band curvature, and to allow the filled particles to democratically sample the whole Brillouin zone, we always work in the flat-band limit of a topological insulator. This corresponds to keeping the single particle eigenstates of $h_2(k)$ but putting the energies of the occupied bands to be an arbitrary energy $\pm E_0$ where $E_0>0$. At the Hamiltonian level, we transform from $h_2(k) = E_-(k) P_-(k) + E_+(k) P_+(k)$ to $h_2^{FB}(k) = -E_0 P_-(k) + E_0 P_+(k)$ where $P_\pm$ are the projectors onto the occupied and unoccupied bands. As such, the energy difference between the valence and conduction bands can be made large without changing the eigenstates of the system. We diagonalize the interaction Hamiltonian directly in the filled band, neglecting the conduction band. This is similar to the Lowest-Landau Level (LLL) projection in the usual fractional auantum Hall effect.  A nice feature of the checkerboard lattice model is that the Hubbard interaction is fixed and has a simple form:
\beq
H_{\text{interaction}} = \sum_{<ij>} n_i n_j
\eneq where $i,j$ are nearest neighbor sites. One can see that the interaction only couples $A$ with $B$ sites. This is not the case for multi-orbital models, in which case there is both an on-site and nearest neighbour interaction. Upon Fourrier transform, and gauge transformation, the interaction reads:
\beq
\frac{1}{N}\sum_{k_{1,2,3,4}} \delta_{\vec{k}_1+\vec{k}_3- \vec{k}_2-\vec{k}_4; \text{mod} 2\pi} V_{k_1, k_2, k_3, k_4} c_{k_1A}^\dagger c_{k_2 A}c_{k_3B}^\dagger c_{k_3 B} \nonumber
\eneq where 
\beq
V_{k_1, k_2, k_3, k_4}  = (1+e^{i(k_{4y} - k_{3y})})(1+e^{-i(k_{4x} - k_{3x})})
\eneq 
In passing, we remark that the present interaction and matrix elements are, in principle, very different from the pseudopotential Hamiltonians that give rise to the fractional quantum Hall effect in the lowest Landau level.

\section{Numerical Procedure and results}

\begin{figure}[tbp]
\includegraphics[width=3.4in]{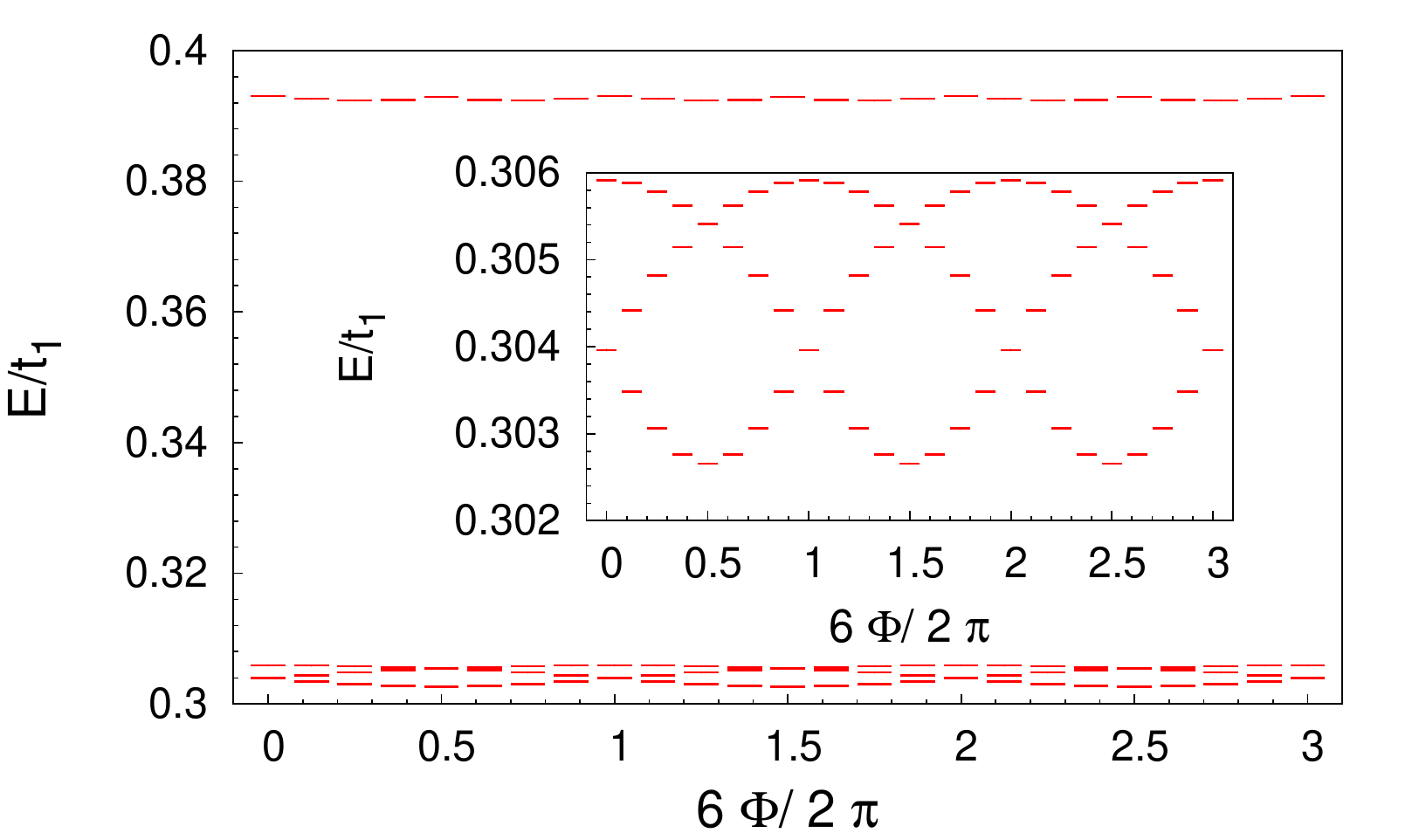}
\caption{Evolution of the $3$-fold degenerate ground state upon flux insertion  along the $y$ direction at $N=10, N_x=5$ and $N_y=6$. The $3$-fold degenerate ground states spectral flow into each other (inset) separated at each point in the flux insertion from the first excited state (the only one of the excited states shown here) which does not exhibit spectral flow with any of the other states.}\label{FluxInsertionGS}
\end{figure}

We now diagonalize this Hamiltonian for $N$ particles in a $N_x \times N_y$ lattice, where, for the ground state we have $N_x \cdot N_y = 3 N$ (we concentrate on the $\nu=1/3$ filling), while for the quasihole excitations we have  $N_x \cdot N_y > 3 N$.  This setup is quite different from the fractional quantum Hall effect on a lattice \cite{kol-1993PhysRevB.48.8890,Moller-2009p184,Moller-2010p828} where the lattice size can be changed while keeping fixed both $N$ and $\nu$. In our case, only the lattice aspect ratio might be tuned in some cases (not all of them are accessible for a given $N$ and $\nu$). All the numerical calculations are performed with $t_2=(2-\sqrt{2})/2 t_1$ as discussed in \cite{sun-2010arXiv1012.5864S}. The total translation operators in the $x,y$ directions commute with both the single and many-body Hamiltonians and hence the eigenstates are indexed by total momentum quantum numbers $(K_x, K_y)$ which are the sum of the momentum quantum numbers of each of the $N$ particles modulo $(N_x, N_y)$. The basis states are $\prod_{i=1}^N \gamma^\dagger_{-, \vec{k_1}} \ldots \gamma^\dagger_{-, \vec{k_N}} \ket{0}$ (we work in the "LLL", and the $\gamma^\dagger_{-,\vec{k}}$'s are the creation operators for a particle of momentum $\vec{k}$ in the valence band). When acting on the basis states, the $c_{\vec{k}, \alpha} = u_{-,\alpha, \vec{k} }\gamma_{-, \vec{k}}$ where $u_{-,\alpha, \vec{k} }$ is the $\alpha = A,B$ component of the eigenstate of the occupied band of $h_2(k)$ or $h_2^{FB}(k)$ (they have identical eigenstates). Diagonalizing directly in the valence band provides for large numerical efficiency. The inversion symmetry of the single particle problem is maintained at the level where the many-body interaction is taken into account. Thus the spectrum has an exact  $(K_x, K_y) \rightarrow (-K_x, -K_y) $ symmetry which can be used as checkup. We first perform all the calculations for the $M=0$ nontrivial Chern insulator, and then drive the fractional Chern insulator to a phase transition by increasing $M$.

\begin{figure}[tbp]
\includegraphics[width=3.5in]{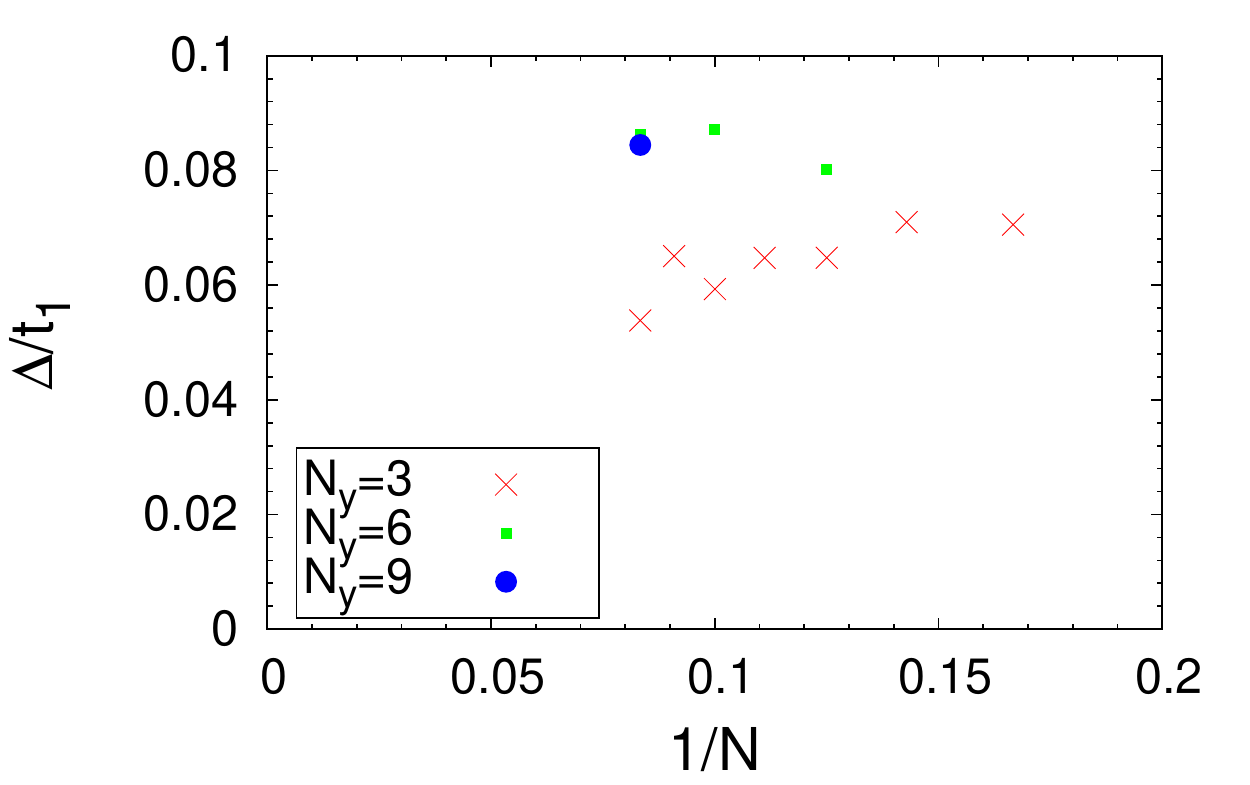}
\caption{Energy gap $\Delta$ for different system sizes and aspect ratio. The gap is defined as the difference between the energy of the first excited state and the highest energy of the $3$-fold ground state manifold. In each case, $N_x=3N/N_y$.}\label{gapscaling}
\end{figure}

\subsection{Degenerate Ground State and its Properties}

In Fig.~\ref{groundenergy} we show the spectrum of the system for several sizes $N=6,10, 12$ and aspect ratios $(N_x, N_y) = (3,6), (5,6), (6,6)$ in the $M=0$ nontrivial Chern insulator case . The choice of aspect ratios will be explained later.  We observe a $3$-fold degenerate ground state at lattice momenta $\{(0,3),(0,3),(0,3)\}$, $\{(0,1)(0,3)(0,5)\}$, $\{(0,0), (0,0) (0,0)\}$ respectively. The principle that determines these momenta will be explained  later. The insertion of flux quantum in either of the $x,y$ directions can be performed by letting each single particle momentum $k_{x,y} \rightarrow k_{x,y} + \Phi/N_{x,y}$ with $\Phi \in [0, 2 \pi]$. For the case $N=10$, Fig.~\ref{FluxInsertionGS} shows that the $3$ degenerate states (which occur at the same lattice momentum and are hence split by the interaction) experience spectral flow into each other (as also previously in Ref.~\cite{neupert-PhysRevLett.106.236804}). In the presence of a gap, $3$-fold degenerate ground states that experience spectral flow in a $\nu=1/3$ fractionally filled Chern insulator  are enough to guarantee the presence of a $\sigma_{xy} = 1/3$ fractional quantum Hall state. Upon the insertion of $3$ fluxes, the ground states move back to the initial configuration, but, as the filled band has Chern number unity, one electron has been transferred across the gap, giving a Hall conductance of $\sigma_{xy} = 1/3$. These arguments have already been presented in \cite{sheng-natcommun.2.389, neupert-PhysRevLett.106.236804} but they are valid only in the presence of a spectral gap, which was not proved in \cite{sheng-natcommun.2.389, neupert-PhysRevLett.106.236804}. In the absence of such a proof, the effects presented in \cite{sheng-natcommun.2.389, neupert-PhysRevLett.106.236804} could also occur in a $1/3$ charge density wave. The existence of a gap in the thermodynamic limit is complicated by commensuration effects on the lattice. Because of the system sizes that can be accessed numerically, we cannot give a quantitative value for the size of the gap in the thermodynamic limit. Our results do, however, strongly suggest that this value should be finite.

The finite-size scaling of the  gaps $\Delta$ is presented in Fig.~\ref{gapscaling}. Several crucial trends are visible. For $N_y=3$, the gap is finite but decreasing as we increase $N$ (or equivalently $N_x= 3N/N_y$). Its decrease does \emph{not} mean that the FQH state is compressible in the thermodynamic limit. Indeed, the thermodynamic limit here does \emph{not} correspond to a $2$-dimensional system: for $N\rightarrow \infty$, if $N_y$ is kept fixed at $3$, we reach the case of a one-dimensional system. This system, even at single particle level, cannot have a Chern number- as discussed later, Chern numbers appear only in the true $2$-dimensional limit. As such, it is expected that for fixed $N_y$, the gap diminishes in the thermodynamic limit. Going to $N_y=6$ shows a \emph{large increase} in the gap from $N_y=3$ exactly because the aspect ratio changed and the system is more two-dimensional. However, if we keep $N_y$ fixed, the gap will also start diminishing as we go to the thermodynamic limit, a fact clearly shown in Fig.~\ref{gapscaling}. When we go to $N_y=9$, the gap increases again. As such, we are confident that the gap remains open and scales to a finite value for $N_x/N_y \rightarrow finite$, $N\rightarrow \infty$

\begin{figure}[tbp]
\includegraphics[width=3.5in]{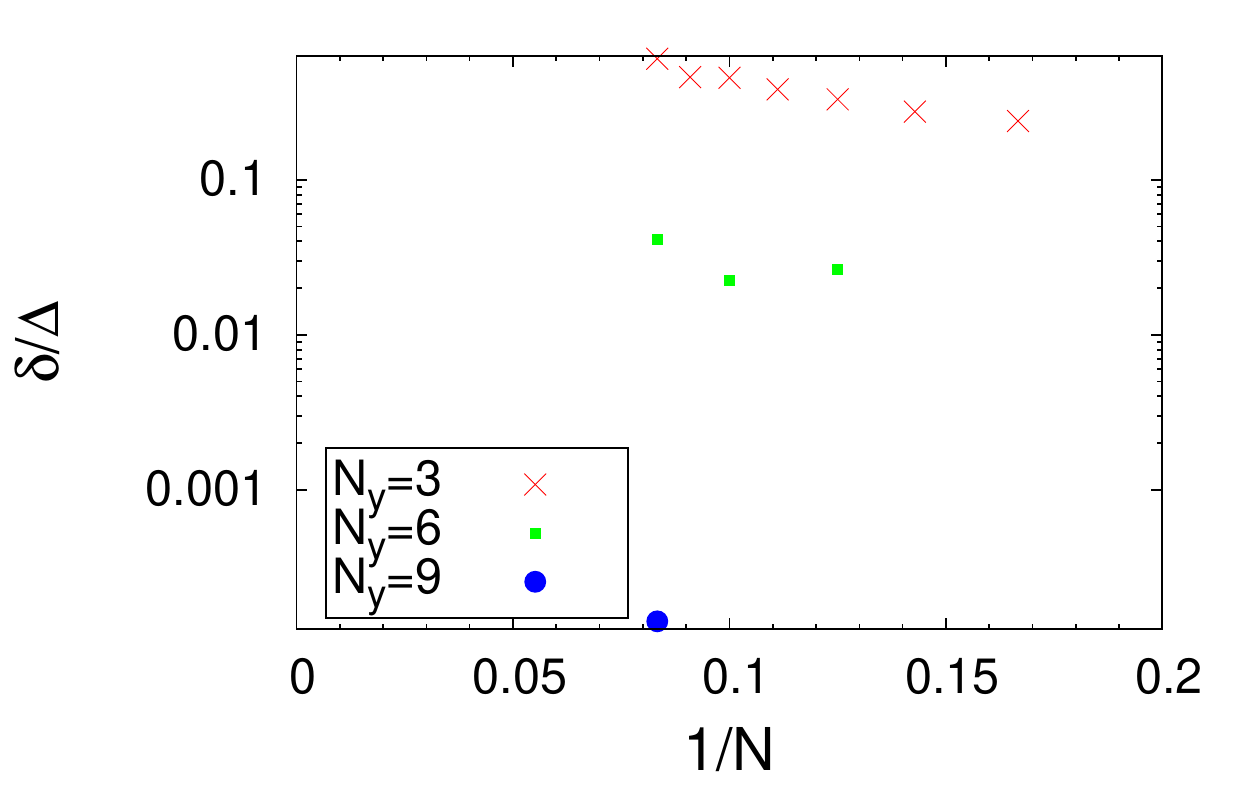}
\caption{Ratio between the energy spread $\delta$ of the $3$-fold ground state manifold and the energy gap $\Delta$ for various system sizes and aspect ratio. The spread is defined as the difference between the highest energy and the lowest energy of the $3$-fold ground state manifold. In each case, $N_x=3N/N_y$.}\label{gapspread}
\end{figure}

The energy spread is plotted in Fig.~\ref{gapspread}. If $E_1, E_2, E_3$ are the energies of the $3$ quasi-degenerate ground states and $E_4$ is the energy of the first state above the lowest $3$ states, then the spread $\delta = E_3-E_1$ while  the gap $\Delta = E_4-E_3$.  Fig.~\ref{gapspread} reveals the degree of degeneracy relative to the size of the gap as a function of $N = N_x \cdot N_y/3$ for fixed $N_y$ . We see that the degeneracy gets better and better for larger sizes.

\begin{figure}[tbp]
\includegraphics[width=3.5in]{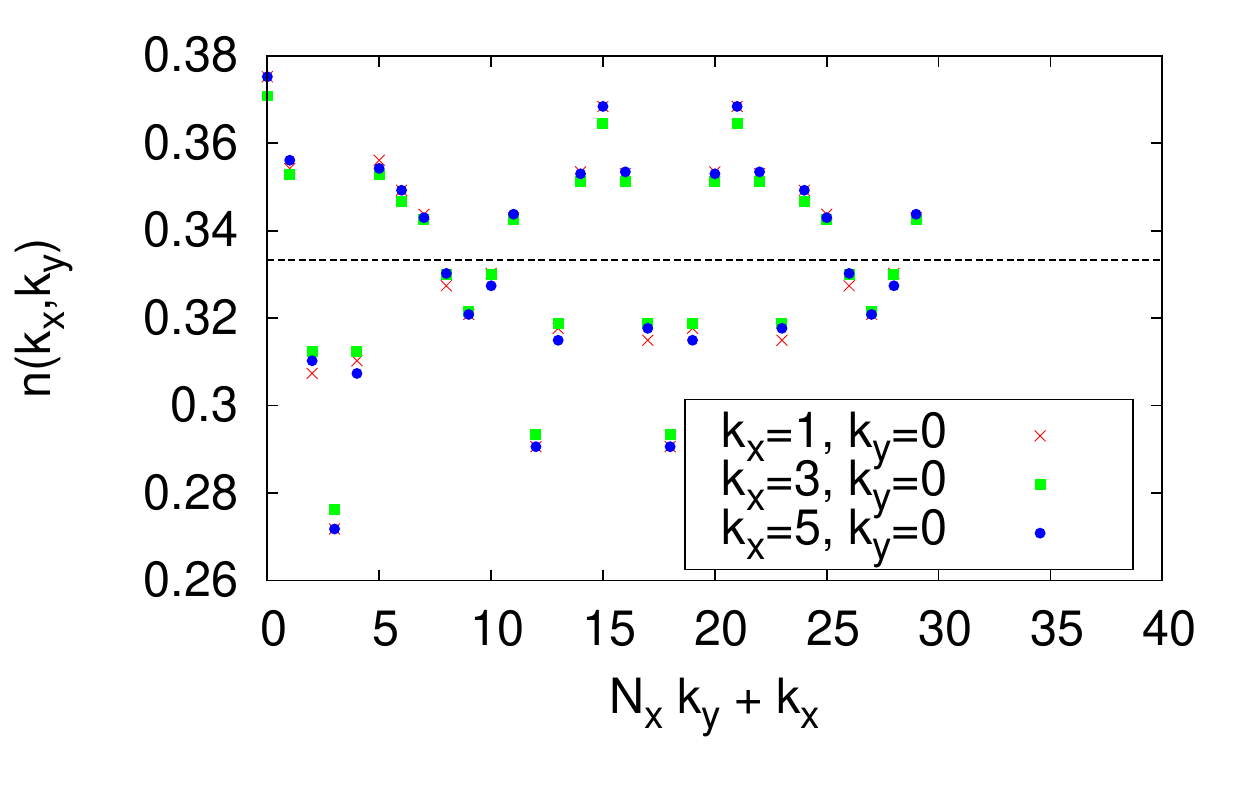}
\caption{  Occupation number of each of the single-particle momentum orbitals for the three $N=10$ ground states (which occur at different total momenta) at $N_x=5, N_y=6$ . Notice that the occupation number is uniform and close to the filling factor $1/3$}\label{momentumspacedensity}
\end{figure}

We have also plotted the momentum-space density of the $3$ degenerate states in the Fractional Chern insulator. In a FQH state on the sphere or on the torus, the occupation  of every angular momentum orbital is constant and equal to $1/3$ (the filling factor) in the thermodynamic limit. In the current case, we expect in finite size deviations from the constant density $1/3$ scenario, due to the fact that the Berry curvature is not uniform in the Brillouin zone. We however, also expect that the deviations not be very large because, in the end, the state must be an incompressible liquid.  The standard deviation of the momentum space density strongly depends on the aspect ratio, typically $\sim 0.1$ for $N_y=3$ and $\sim 0.02$ for $N_y=6$ (almost independent from $N$). The results are presented in Fig.~\ref{momentumspacedensity}. We observe that the occupation numbers of each of the momenta of the $3$ degenerate ground states for the $N=10$ particle state are close to $1/3$, as expected for an incompressible liquid at this filling. As the $3$ ground states occur at different momenta, the fact that each of the ground states has a uniform $n(k)$ close to $1/3$ for any value of $k$ removes the possibility that these states are charge-density waves.

\subsection{Quasihole excitations}

To iron out any doubt about the existence of a FQH state, we now investigate the quasihole excitations of the system. The excitations of a FQH state are very specific in the sense that they carry fractional $1/3$ charge and exhibit fractional $1/3$ statistics. This would not happen in a charge-density wave state. The quasihole excitations arise by inserting a flux in the FQH state. These properties endow the excitations with a state counting specific to the FQH universality class at a certain filling independent of the specific Hamiltonian to be diagonalized or even of the form of the model wavefunction. In the current case, it is expected that the actual form of the ground state wavefunction is \emph{not} of a simple known Laughlin form. However, it is also expected that the FQH at filling $1/3$ is the Abelian FQH $1/3$ state, whose universal properties (after integrating out the fermions) are described by the topological Abelian Chern-Simons field theory. As such, its universal quasihole counting (even in finite size) should be identical to that of the usual Laughlin quasihole states. In fact, as our interaction represents a generic Hamiltonian (similar to the Coulomb Hamiltonian in the FQH case), all we can hope for is that the set of quasiholes is separated by a finite gap from higher energy non-universal excitations that are not described by the topological field theory.

The FQH $\nu=1/3$ quasihole counting is identical to that of particles with Haldane $1/3$ statistics in both the thermodynamic limit and in the finite size systems. This counting can be directly obtained from the topological Chern-Simons field theory and the assumption of bulk-edge correspondence: In a system with the boundary, the Abelian $1/\nu$ Chern-Simons field theory is not gauge-invariant - it is missing an edge piece which is a boson at compactification radius $\sqrt{1/\nu}$. The modes of this boson are the edge modes, a gapless conformal field theory whose excitations, when placed in a finite-size box exhibit Haldane $1/\nu$ statistics: Every particle must be separated from another particle by at least $1/\nu$ orbitals. This is called $(1,1/\nu)$ generalized Pauli principle, which will be presented in the next section. However, the bulk-edge correspondence renders this counting also the same as that of the quasiholes in the bulk of the FQH state, and represents a hallmark of the  $\nu=1/3$ FQH state. 
\begin{figure}[tbp]
\includegraphics[width=3.5in]{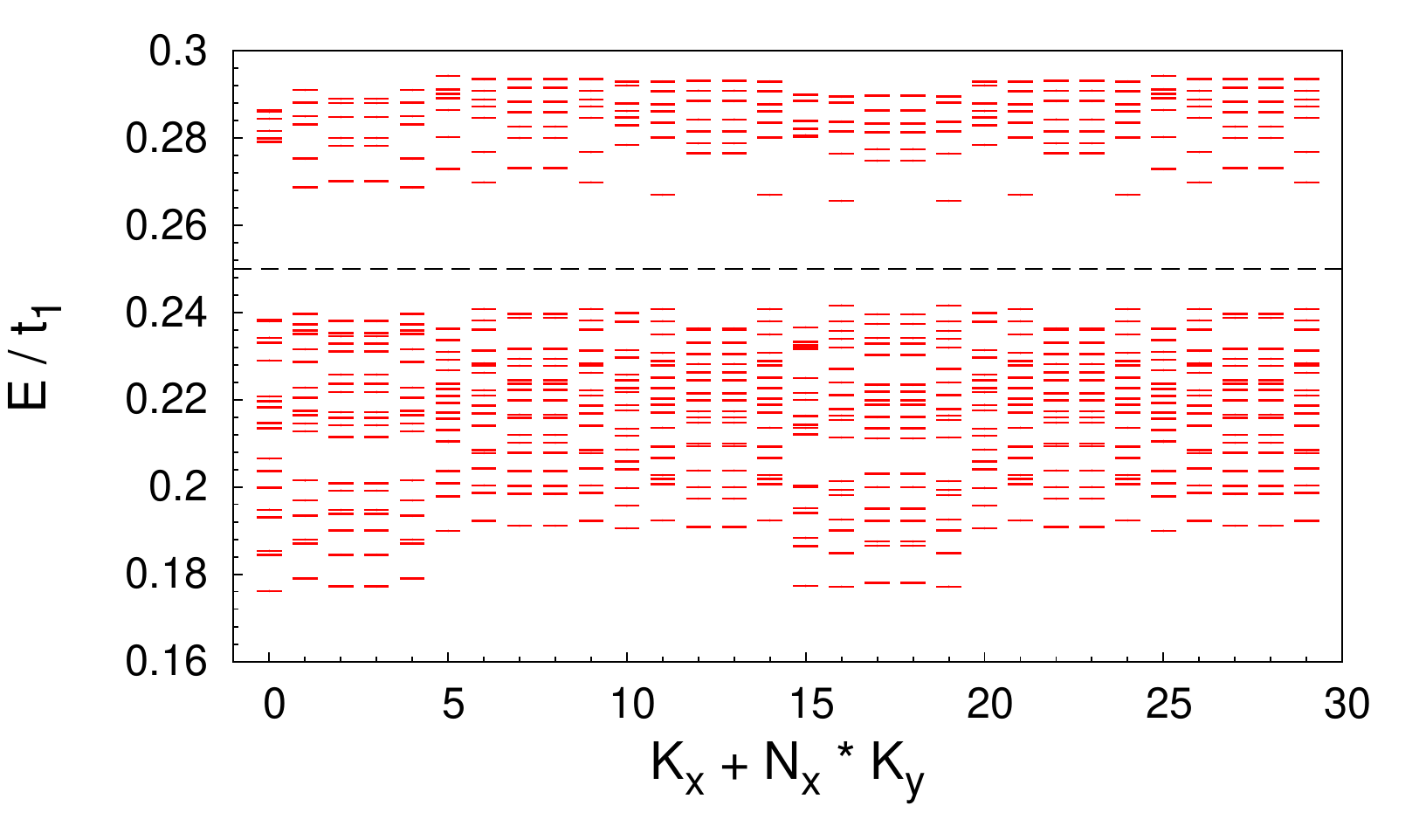}
\caption{Low energy spectrum for $N=9, N_x=5, N_y=6$. The number of states below the dashed line is 19 in sectors where $K_y \mod 3 =0$, 18 in other sectors, as expected from the analytical results. The total counting matches the one expected for Laughlin quasihole states. The counting for each momentum sector is given by the generalized Pauli principle  and two dimensional to one dimensional unfolding presented in the next section. }\label{spectrumqh9}
\end{figure}

We now compare such a counting with that obtained from numerics. For the model FQH Hamiltonians such as the Haldane pseudopotentials, the quasiholes are just the degenerate ground states of the system when fluxes are added to the system. In this sense, the ground state is just the highest density state of a set of other degenerate states which occur at higher fluxes and which describe the quasiholes. As we cannot add unit fluxes to a ground state with a set aspect ratio and still keep the translational symmetry intact, we choose to analyze quasiholes by either increasing $N_{x} \rightarrow N_{x} +1$ (or $N_{y} \rightarrow N_{y} +1$) or by  removing particles from a ground state and keeping the aspect ratio fixed  - hence our system sizes contain at least $3$ fluxes above the ground state. A generic example is that shown in Fig.~\ref{spectrumqh9}, in which $N=9$ particles reside in $N_x\times N_y = 5\times 6$ orbitals, $3$ fluxes more than the ground state at filling $1/3$. We diagonalize the interacting Hamiltonian for this number of particles and observe (Fig.~\ref{spectrumqh9}) that the spectrum splits into two parts separated by a clearly visible and unambiguous energy gap. The counting of states \emph{below} the gap equals the counting of quasiholes of a $1/3$ fractional quantum Hall state of $9$ particles in $30$ orbitals (explained in the next section). Even more, the counting of states per momentum sector also matches that of the generalized Pauli principle for the quasihole states \cite{bernevig-PhysRevLett.100.246802}  previously obtained in the FQH context. This  will also be explained in the next section.

\begin{figure}[tbp]
\includegraphics[width=3.5in]{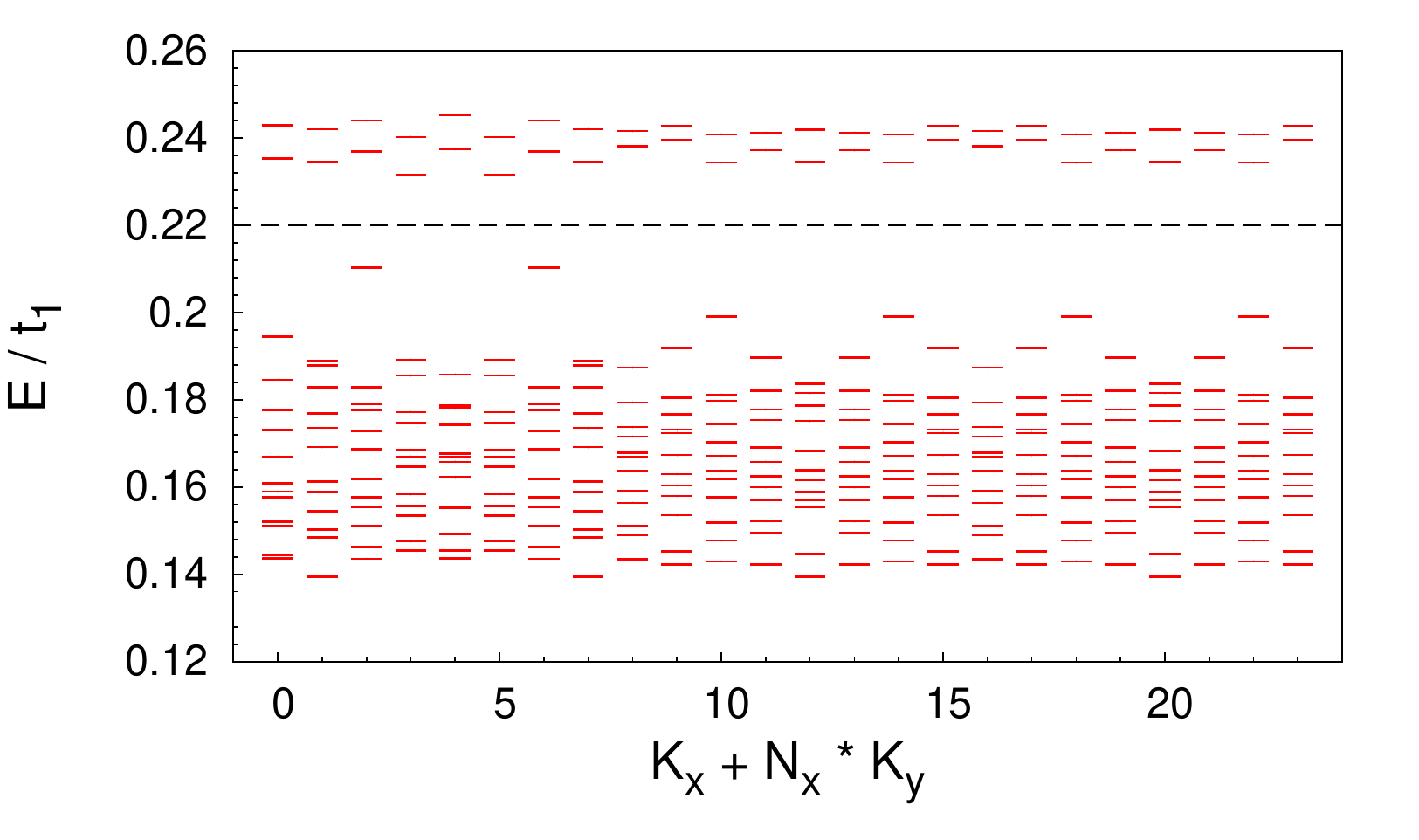}
\caption{Low energy spectrum for $N=7, N_x=8, N_y=3$. The number of states below the dashed line is 12 in all sectors and matches the number of expected quasihole states given by the generalized Pauli principle.}\label{spectrumqh7}
\end{figure}

\begin{figure}[tbp]
\includegraphics[width=3.5in]{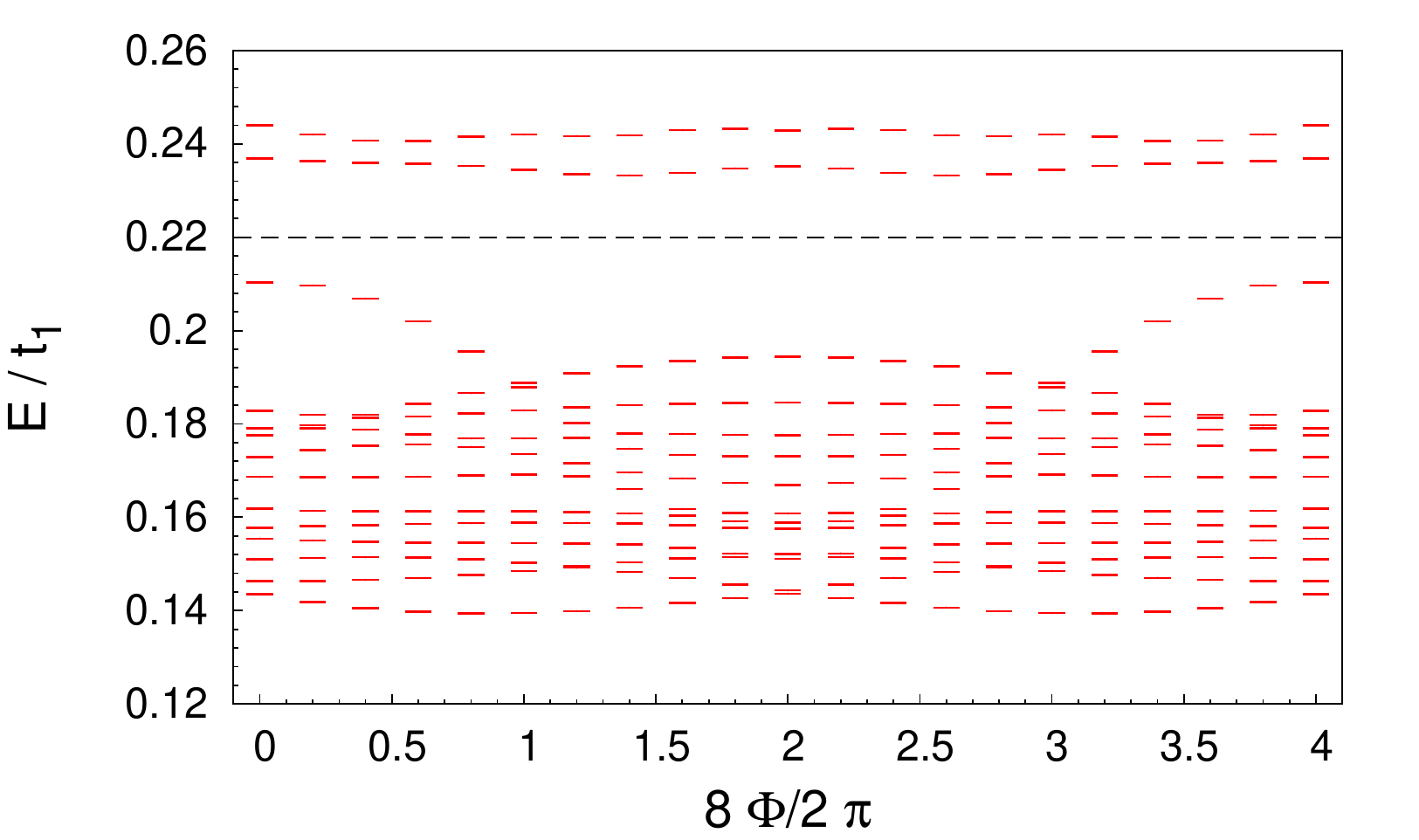}
\caption{Evolution of the low energy spectrum for $N=7, N_x=8, N_y=3$ at $K_x=2, K_y=0$ upon flux insertion  along the $x$ direction. The number of states below the dashed line is 12 for any value of $\Phi$ and matches the number of expected quasihole states given by the generalizaed Pauli principle.}\label{spectrumflowqh7}
\end{figure}

While for $N=9$, $N_x \times N_y = 5 \times 6$ the gap that separates the quasihole manifold from the spurious states is clear, this is not always the case. In Fig.~\ref{spectrumqh7} we plot the spectrum of the $N=7$, $N_x \times N_y = 8 \times 3$ problem:  for most momentum sectors we find a clear gap, visible by eye, between a low-energy quasi-hole manifold and that of spurious, high-energy states. However, for the momenta $(K_x, K_y)=(2,0)$ (and for its inversion symmetric $(6,0)$) we see one state (at energy around $0.21$) which lies in the gap.  It is a-priori unclear whether, just based on the numerical information, this state belongs to the quasihole manifold or not (the analytic counting obtained in the next section would indeed tell us that it does belong to the quasihole manifold). To find this out, we look at the level spectral flow of the states  at $(K_x, K_y)=(2,0)$ upon the flux insertion $k_{x} \rightarrow k_{x} + \Phi/N_{x}$ in  Fig.~\ref{spectrumflowqh7}, where we insert flux in the $x$ direction.  We note that the state at energy $0.21$ exhibits level spectral flow with the states below it, and hence it belongs to the quasihole subspace. Note that the first two excited states (around energy $0.24$) do not mix with the quasihole manifold upon flux insertion and even exhibit level repulsion between themselves. For example, when $\Phi= \pi/8$ the lower energy manifold is clearly separated from the spurious, high-energy one. The flux insertion tells us that the state which, in the periodic boundary condition case was at energy $0.21$ indeed belongs to the quasihole manifold.

\begin{figure}[tbp]
\includegraphics[width=3.5in]{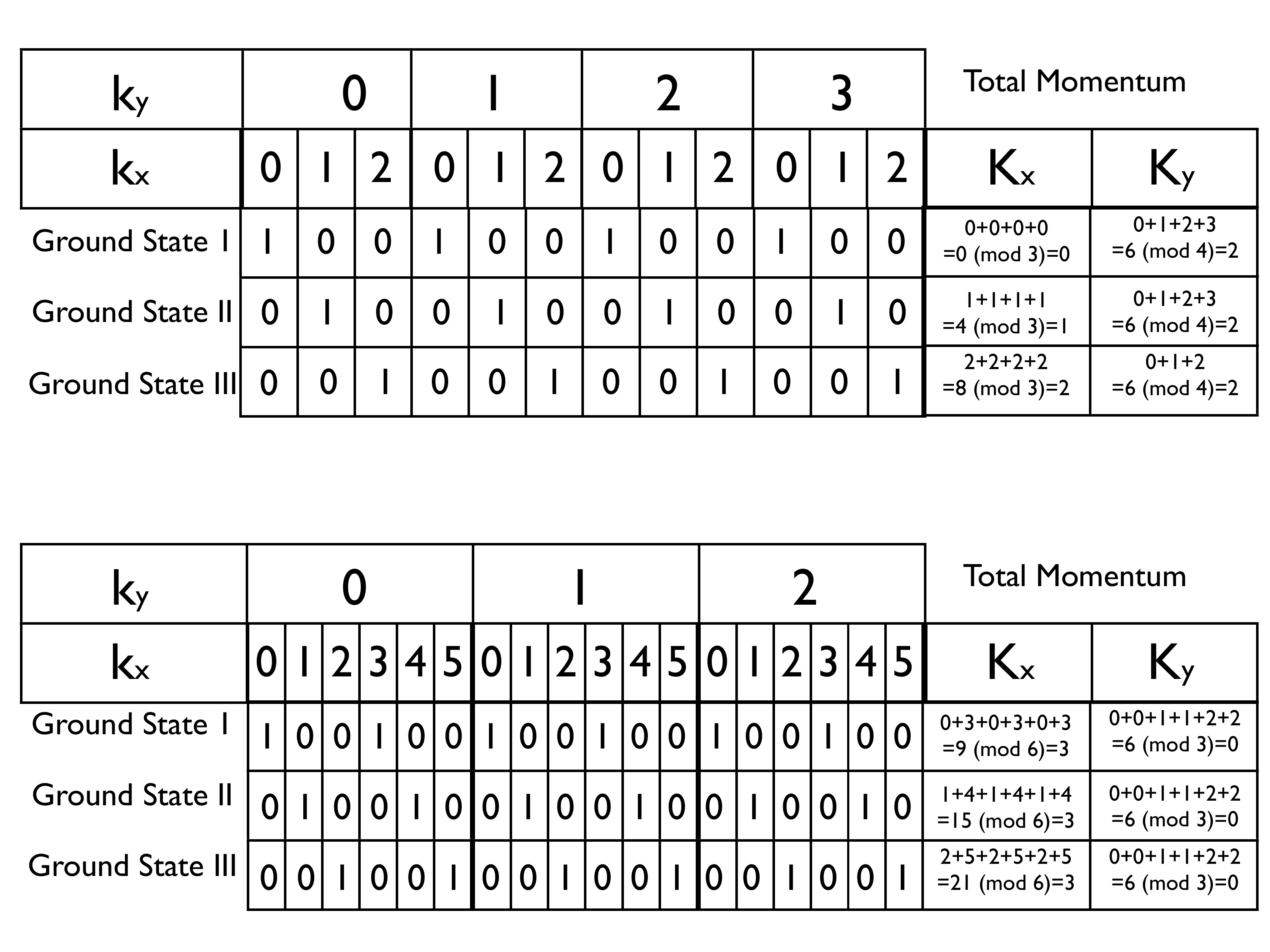}
\caption{Example of total momentum counting for the $3$-fold degenerate ground states of the $N=4$ (top) and $N=6$ (bottom) problem. Every time the number of particles $N$ is a multiple of both $k_x, k_y$, the $3$-degenerate ground states occur at the same total momentum and are expected to be split by the interaction for finite-size samples. Notice that for the $N=4$ problem, the total momenta at which the ground state occurs are related by inversion symmetry, as $(1,2)= (-2,-2)$ mod $(3,4)$  }\label{ground statemomentum1}
\end{figure}

A similar analysis has been made on several other system sizes and aspect ratios for quasiholes close to the $1/3$ filling.  In the absence of a pseudo-potential Hamiltonian and since we work on a lattice, several factors must be taken into account when looking at quasiholes. First, as we are diagonalizing a generic Hubbard-type Hamiltonian, we must make sure that we are still close to filling $1/3$. Taking out too many particles (creating too many quasiholes) could potentially lead us to another filling factor and diminish or even erase the gap between the low-energy quasiholes and the spurious high energy states. Since we work on the lattice, care should be taken that the single-particle system is not one-dimensional - the aspect ration should be roughly balanced. This will be discussed later. In most cases when these two conditions are satisfied, the gap between the quasihole manifold and the upper spurious states is visible. The total number of states (in all momentum sectors) \emph{below} the spectral gap matches exactly the number of states of the Laughlin quasiholes in the regular FQH effect on the torus of $N$ particles  and $N_x\cdot N_y$ number of orbitals. This is the number of $(1,3)$-admissible partitions on the torus (as defined below, see Figs.~\ref{flatdeform} and \ref{ground statemomentum1}) that we can write down out of $N$ particles in $N_x \cdot N_y$ orbitals.  For $N=8$ and $N_x, N_y = 5,6$ ,respectively, we obtain a total of $6435$ states  while for $N=9$ we obtain 550 states by counting these partitions. This number is identical to the numerical data (see Fig.~\ref{spectrumqh9} and \ref{spectrumqh7}). We have checked a series of sizes  up to $N=10$ and have confirmed in each case the existence of a large gap between a high energy nonuniversal sector and a low-energy quasihole sector whose counting matches exactly that of the Laughlin quasiholes obtained by counting $(1,3)$-admissible partitions. All these quasihole states exhibit spectral flow within themselves. This proves that the observed state is an Abelian fractional quantum Hall state with $\sigma_{xy}=1/3$, as it has excitations obeying Haldane $1/3$ statistics and having the counting of a $U(1)$ boson. The overall quasihole counting matches the $(1,3)$ generalized Pauli principle. But in most cases, this statement is also valid for the counting per each momentum sector. The $(1,3)$ generalized Pauli principle is generic of the Laughlin FQH state at filling $1/3$ and will be presented below.

\section{Heuristic Counting rule: Pauli Principle and Crystal Momenta}

\begin{figure}[tbp]
\includegraphics[width=3.5in]{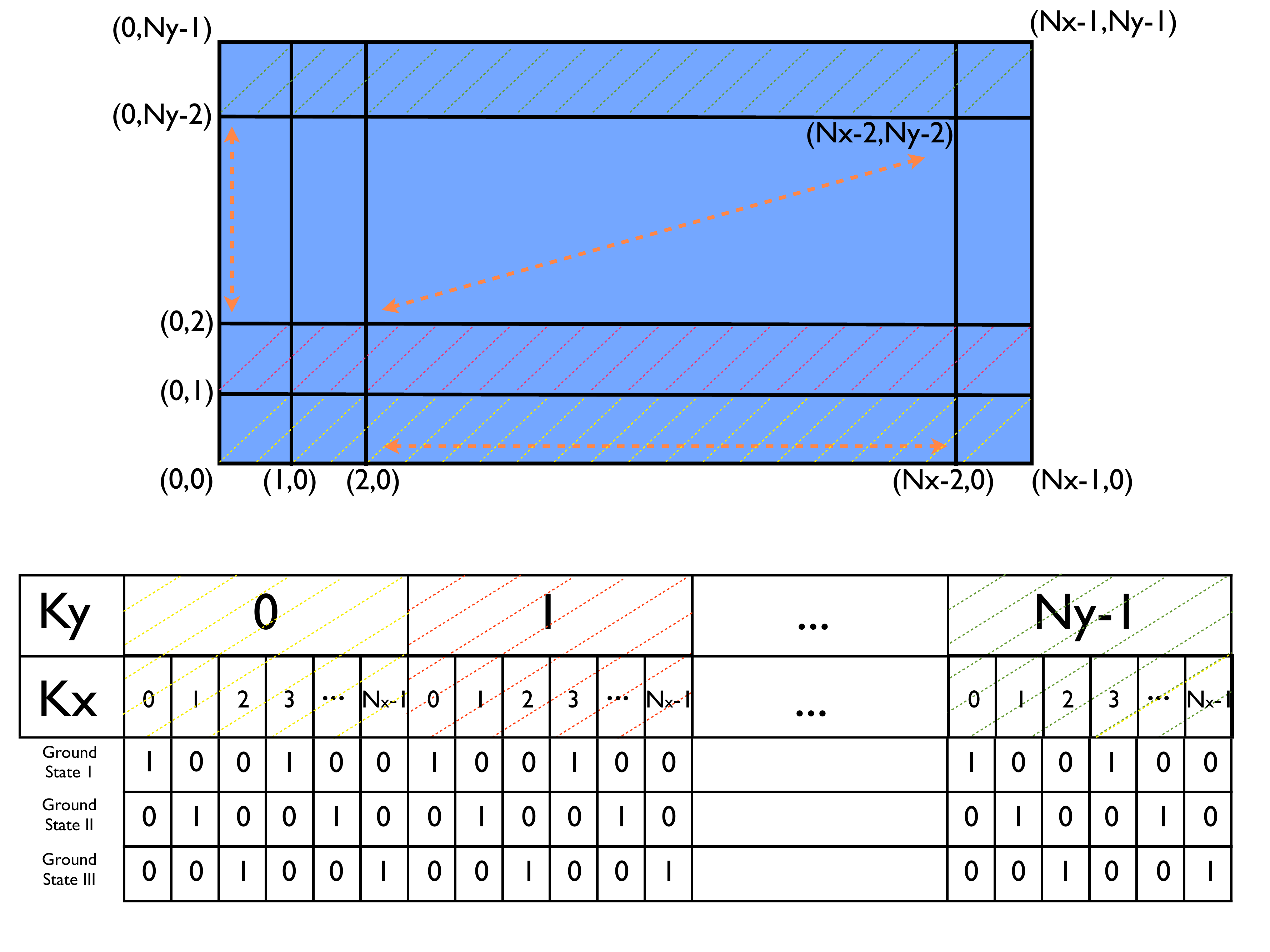}
\caption{Momentum of the $3$-fold degenerate ground state for the fractional Chern insulator at filling $\nu=1/3$. The $2$-dimensional lattice of momenta $k_x, k_y$ with $k_x \in [0, N_x-1] $ and  $k_y \in [0, N_y-1] $ (in units of $2 \pi/N_x, 2\pi/N_y$ respectively) where we have taken $N_x$ to be a multiple of $3$ (since $N_x \cdot N_y = 3N$, this can always be done) is unfolded in a $1$-d lattice. The total momentum of the ground states is then the same as the momentum of the three $(1,3)$-admissible partitions possible for the number of orbitals $N_x \cdot N_y$ and number of particles $N =N_x \cdot N_y/3$}\label{flatdeform}
\end{figure}

We now present a counting rule for the total lattice momentum at which the degenerate ground states occur. A more detailed version will be developed in an upcoming paper\cite{bern-reg-prep}. The $2$-dimensional lattice of sites $N_x \times N_y$ with $k_x \in[0, N_x-1]$ and $k_y \in [0, N_y-1]$ can be folded into a $1$-dimensional lattice of orbitals of momentum $\lambda = N_x k_y + k_x$ by placing the $x$ momenta range one after each other for the different values of the $y$-momenta, as per Fig.~\ref{flatdeform}. This then represents a map between the momentum of the $i$'th particle orbital $k_{x,i}, k_{y,i}$ and the $1$-dimensional momentum of the $i$'th orbital $\lambda_i \in [0,N_x\cdot N_y -1]$ . The $\lambda_i$'s then form a partition $\lambda =[\lambda_1, \ldots \lambda_N]$ which we order from highest to lowest. 

This partition is then similar to the orbital momenta of the particles of the usual FQH on the torus. On such geometry and in the lowest Landau level, the single-particle orbitals can be written as
\begin{eqnarray}
\phi^{\rm FQH}_{\lambda}(x,y) = \sum_{m \in {\mathbb{Z}}}&& e^{\frac{2 \pi }{L_y}(\lambda+ m N_\phi)(x+ i y)}\\\nonumber
&& \times e^{-\frac{x^2}{2 l_b^2}}e^{-\frac{1}{2} \left(\frac{2 \pi l_b}{L_y} \right)^2(\lambda + m N_\phi)^2}
\end{eqnarray}
In the above, we have picked the Landau gauge $\vec{A} = B x \vec{e}_y$. $l_b$ is the magnetic length, $N_\phi$ is the number of flux quanta and $L_y$ the system size along $y$. With this gauge choice, $\lambda$ is the orbital momentum along the $y$ direction and is such that $\lambda \in [0, N_\phi - 1]$. Setting $N_\phi = N_x \cdot N_y$, we can formally relate the FQH momentum $\lambda$ and the linearized momentum of the FCI $\lambda = N_x k_y + k_x$.

Using this analogy between FQH on the torus and FCI, we now index each many-body state by  a single partition (called "root" partition \cite{bernevig-PhysRevLett.100.246802}) that satisfies a generalized Pauli principle (also called an "admissibility condition") which does not allow the existence of more than $1$ particle in $3$ consecutive orbitals. In partition notation, this reads $\lambda_i - \lambda_{i+1} \ge 3$. Because of the periodicity of the torus, we must also make sure that the first and last particles are separated by at least $3$ orbitals, which reads: $\lambda_N + N_x\cdot N_y -1 - \lambda_1 \ge 3$. At the ground state filling $N= N_x \cdot N_y/3$ (we pick  $N_x$ a multiple of $3$) there are only  three such partitions that correspond to the occupation numbers $100100100 \ldots 100100$ ($K_x,K_y =N_x(N_x-3)/6 ,N(N_y-1)/2$ mod $(N_x, N_y)$), $010010010 \ldots 010010$ ($K_x,K_y =N_x(N_x-3)/6+ N ,N(N_y-1)/2$ mod $(N_x, N_y)$), and $00100100100 \ldots 001001$ ($K_x,K_y =N_x(N_x-3)/6 + 2N ,N(N_y-1)/2$ mod $(N_x, N_y)$), of the orbitals of momentum $\lambda_1, \lambda_2, \ldots \lambda_N$. As an empirical observation, it was noted in \cite{sheng-natcommun.2.389} that the degenerate ground states appear at momenta related by a translation with \emph{the same} number in both $x,y$ directions. This is an immediate corollary of the counting principle presented above.  In the present model, due to inversion symmetry, at least two of the ground states are exactly degenerate. The principle for finding the ground state momentum reminds one of the root partitions used in the usual FQH effect \cite{bernevig-PhysRevLett.100.246802} or in the thin-torus limit of the FQH states \cite{bergholtz-06prb081308}.

At this point, a discussion of the dependence of the physics on the aspect ratio of the problem is necessary. In cases of finite-size systems, the exact diagonalization data for the momentum of the $3$ quasi-degenerate ground states matches the momentum counting presented above for aspect ratios for which the sample is two-dimensional. The Chern insulator problem has the particular property that the filled band only has a Chern number (Hall conductance) equal to unity in the case of  $N_x/N_y \rightarrow \text{finite}$. For example, if we let $N_y=1$ and $N_x \rightarrow \infty$, the Chern number of this Hamiltonian will not be equal to unity. The Chern number is equal to $\int d k_x d k_y Tr\left( P [\partial_{k_x} P, \partial_{k_y} P]\right) / (2 \pi)^2$ and hence $N_y$ cannot take small values such as $1,2$ in order for the derivatives of the Chern number to be well defined. The Brillouin zone mesh of momenta must truly be two-dimensional in order for the filled band to have an integer Chern number. We hence expect that the interacting problem also be sensitive to the aspect ratio when one of the dimensions becomes much smaller than the other (of course in the thermodynamic limit, as long as $N_x/N_y$ remains finite, we expect the degeneracy of the FQH states to be independent of the aspect ratio). In fact, the interacting problem has also a different dependence on the aspect ratio. To see this, one can look at the extreme case of $N_y=1$, $N_x= 3N$. In this case, we are indeed solving a one-dimensional problem of orbitals $k_x=0, \ldots, 3N-1$ - extremely similar to that of the LLL orbitals on the torus -  but with interacting Hamiltonian that comes from a $2D$ Hubbard interaction. This Hamiltonian is much different from the Haldane pseudopotential Hamiltonian which involves Clebsch-Gordon coefficients on the sphere and Jacobi theta functions on the torus. As such, the problem is physically distinct at the interacting level, and for these skewed aspect ratios we do not expect FQH states in the Chern insulator problem. Indeed, in the finite size diagonalizations performed below, we will always aim to choose the most symmetric aspect ratio available. As observed in Fig.~\ref{gapscaling} and \ref{gapspread}, the cleanness of the results will depend on this choice, as it is natural for these finite sizes.

\begin{figure}[tbp]
\includegraphics[width=3.5in]{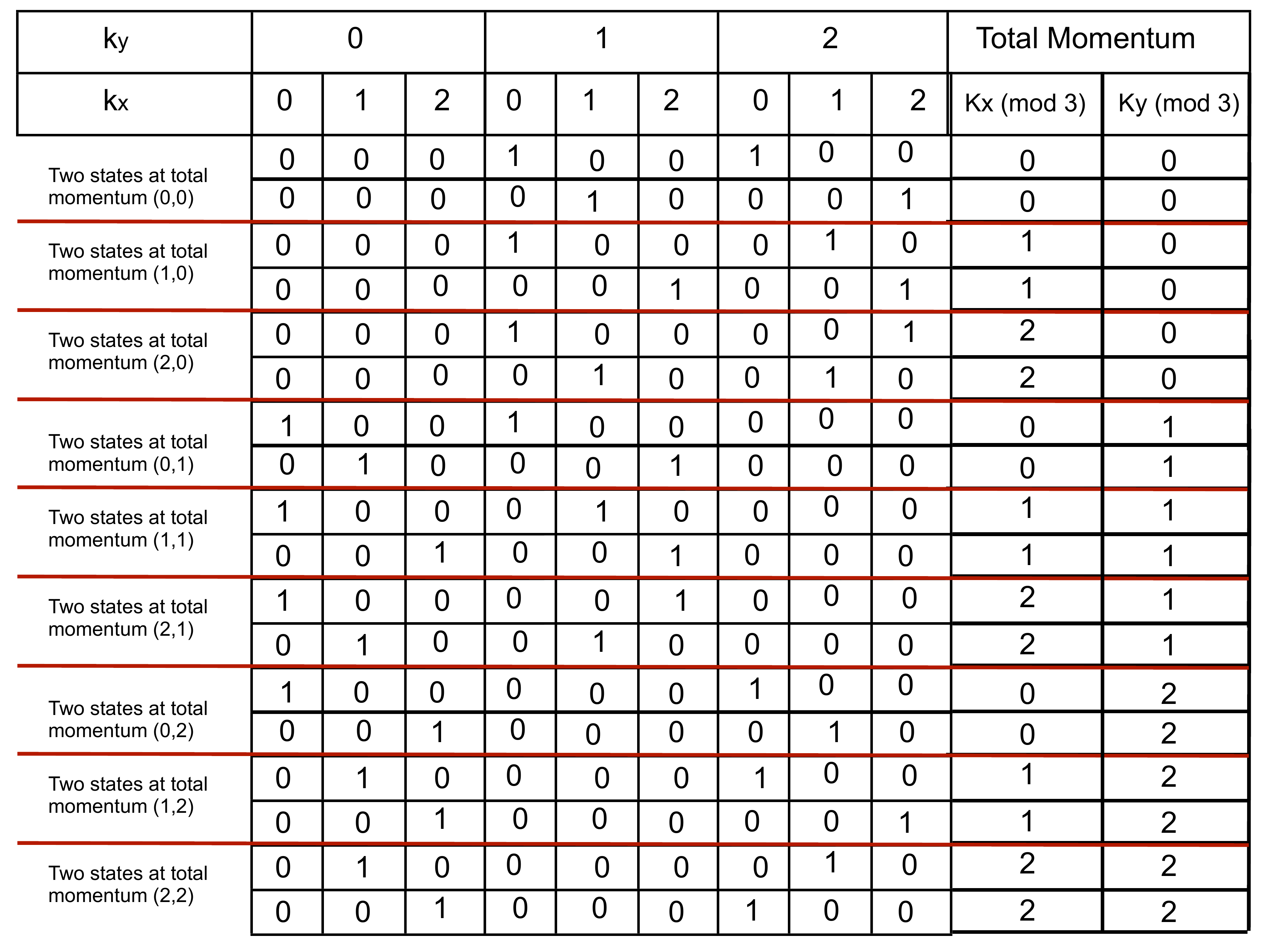}
\caption{Example of total momentum counting for the quasiholes of the $N=2$ $N_x=N_y=3$  problem.}\label{quasiholecounting}
\end{figure}

The counting of quasiholes per momentum sector is slightly more complicated, and we have not found the full counting rule per momentum sector.  The map between the two-dimensional momentum numbers and the one-dimensional orbital partition that allowed us to find the momentum of the ground states cannot always work for the quasihole states. It sometimes violates the inversion symmetry (see the Appendix for such cases) and thus it sometimes does not match the numerical results. Nevertheless, the total number of quasiholes, irrespective of the momentum sectors, always matches the counting of partitions. This matching of the exact diagonalization states and the $1.3$ - admissible partitions is a rather remarkable demonstration of the $\nu=1/3$ character of the states.  Even more interestingly,  the Pauli principle and the map between the two-dimensional momentum numbers and the one-dimensional orbital partition works for \emph{each momentum sector (not only for the total number of quasiholes)} in the great majority of the cases studied.  As an example, we present in Fig.~\ref{quasiholecounting}  the generalized Pauli counting principle at work for $N=2$, $N_x=N_y=3$.

\section{Entanglement spectrum}

We now turn back to the $3$-fold degenerate ground state of the system. There are several attempts to obtain an analytical expression of the Laughlin state for the fractional Chern insulators \cite{vaezi-2011arXiv1105.0406V,qi-PhysRevLett.107.126803}. Unfortunately, none of them can be used in our case, so we cannot perform any wavefunction overlap calculations to compare their analytical expression with our system ground state. Nevertheless, we can still show that the ground state contains by itself, information about the Abelian fractional $1/3$ character of the excitation spectrum. This is actually a far better probe than an overlap, since the $1/3$ character is the feature, not the analytical expression. To do this, we use the recently developed entanglement spectrum \cite{li-08prl010504,sterdyniak-PhysRevLett.106.100405} which for a single nondegenerated ground state $\ket{\Psi}$ can be defined through the Schmidt decomposition of $\ket{\Psi}$ in two regions $A$, $B$ (not necessarily spatial):

\begin{equation}
\ket{\Psi}=\sum_i e^{-\xi_i / 2} \ket{\Psi^A_i} \otimes \ket{\Psi^B_i}
\label{schmidt}
\end{equation}
\noindent where $\braket{\Psi^A_i}{\Psi^A_j}=\braket{\Psi^B_i}{\Psi^B_j}=\delta_{i,j}$. The $\exp(-\xi_i)$ and $ \ket{\Psi^A_i}$ are the eigenvalues and eigenstates of the reduced density matrix,  $\rho_A={\rm Tr}_B \rho$, where $\rho=\ket{\Psi}\bra{\Psi}$ is the total density matrix. There is no generalization of (\ref{schmidt}) to degenerate ground states. Still the definition of the entanglement spectrum through the reduced density matrix can be extended to that case. While several schemes can be proposed, it has been observed in \cite{sterdyniak-PhysRevLett.106.100405} that the incoherent summation over the degenerate ground states $\rho=\frac{1}{3}\sum_{i} \ket{\Psi_i}\bra{\Psi_i}$ is good candidate for this generalization. This combination builds a density matrix which commutes with the total translation operators which is a desired feature to sort the $\xi_i$ with respect to the momentum quantum numbers.

Depending on the space where the system is split into $A$ and $B$, be it real, momentum or particle space, different aspects of the system excitations will be revealed through the ES. It was proven that if the regions $A$, $B$ are regions of particles\cite{sterdyniak-PhysRevLett.106.100405}, the particle entanglement spectrum hence obtained by tracing over the positions of a set of $B$ particles gives information about the number of quasiholes of the system of $N_A$ particles and the number of orbitals identical to that of the untraced system. In the case of the usual FQH, the particle entanglement spectrum of a model state contains an identical number of levels (i.e., the number of non zero eigenvalues) as those of the quasiholes.  Thus the counting of non zero eigenvalues does not suffer from the sometimes uncontrolled finite-size effects, as observed with state decomposition in the momentum space \cite{li-08prl010504,hermanns-2010arXiv1009.4199H}. For this reason we will use state decomposition in the particle space as a probe for the fractional Chern insulator.

Away from the model states, like the Coulomb ground state, the ES may exhibits an entanglement gap \cite{li-08prl010504,thomale-10pr180502}. It separates a low-energy structure with perfect quasihole counting and and a high entanglement energy nonuniversal part. But a clear and significant gap is not always observed, even for the $\nu=1/3$ Coulomb state.

\begin{figure}[tbp]
\includegraphics[width=3.5in]{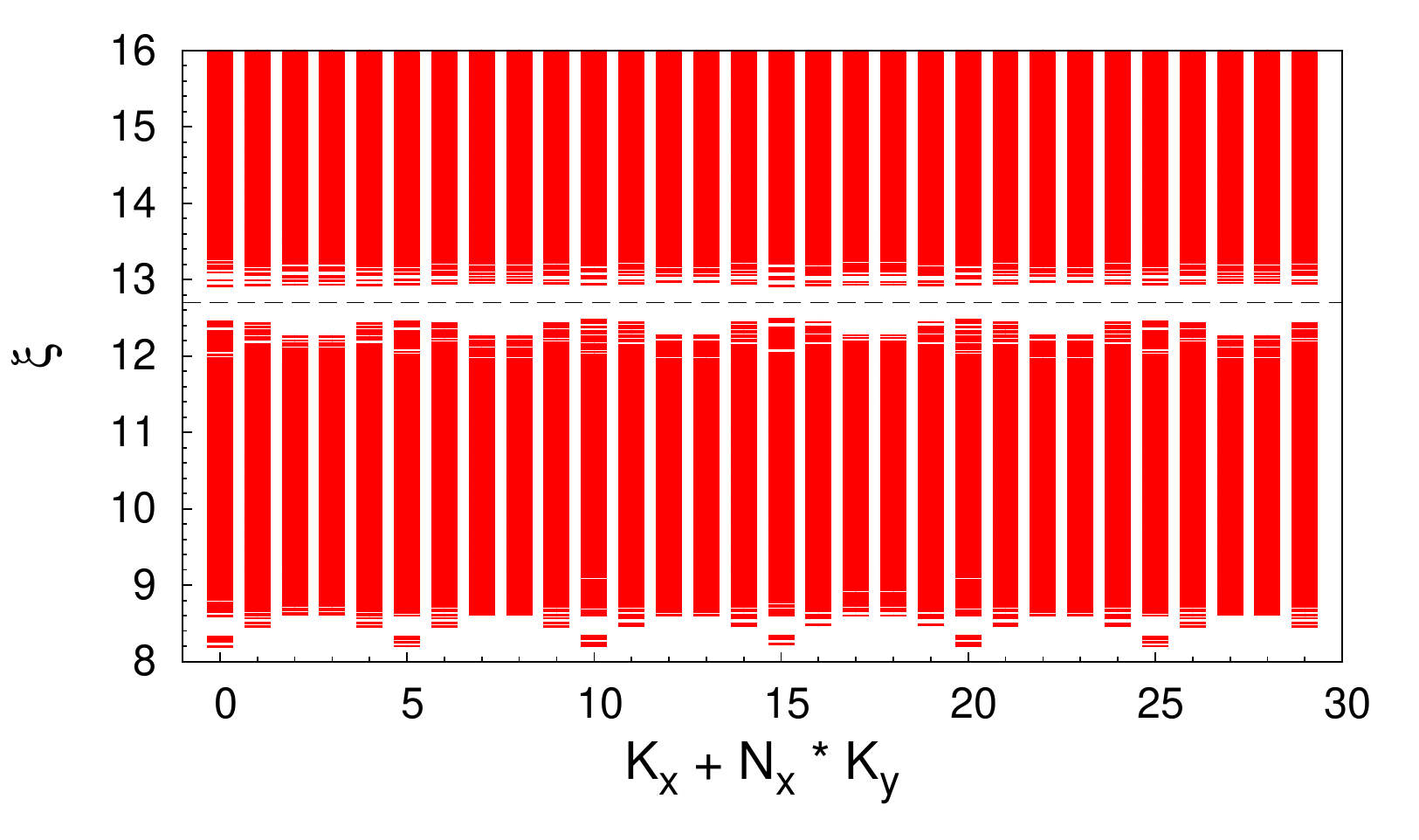}
\caption{Particle entanglement spectrum for $N=10, N_x=5, N_y=6$ and $N_A=5$. The number of states below the dashed line is 776 in all the $k_x=0$, and 775 in the other sectors. This is in agreement with the $(1,3)$ Pauli principle for every momentum sector.}\label{entspec10}
\end{figure}

\begin{figure}[tbp]
\includegraphics[width=3.5in]{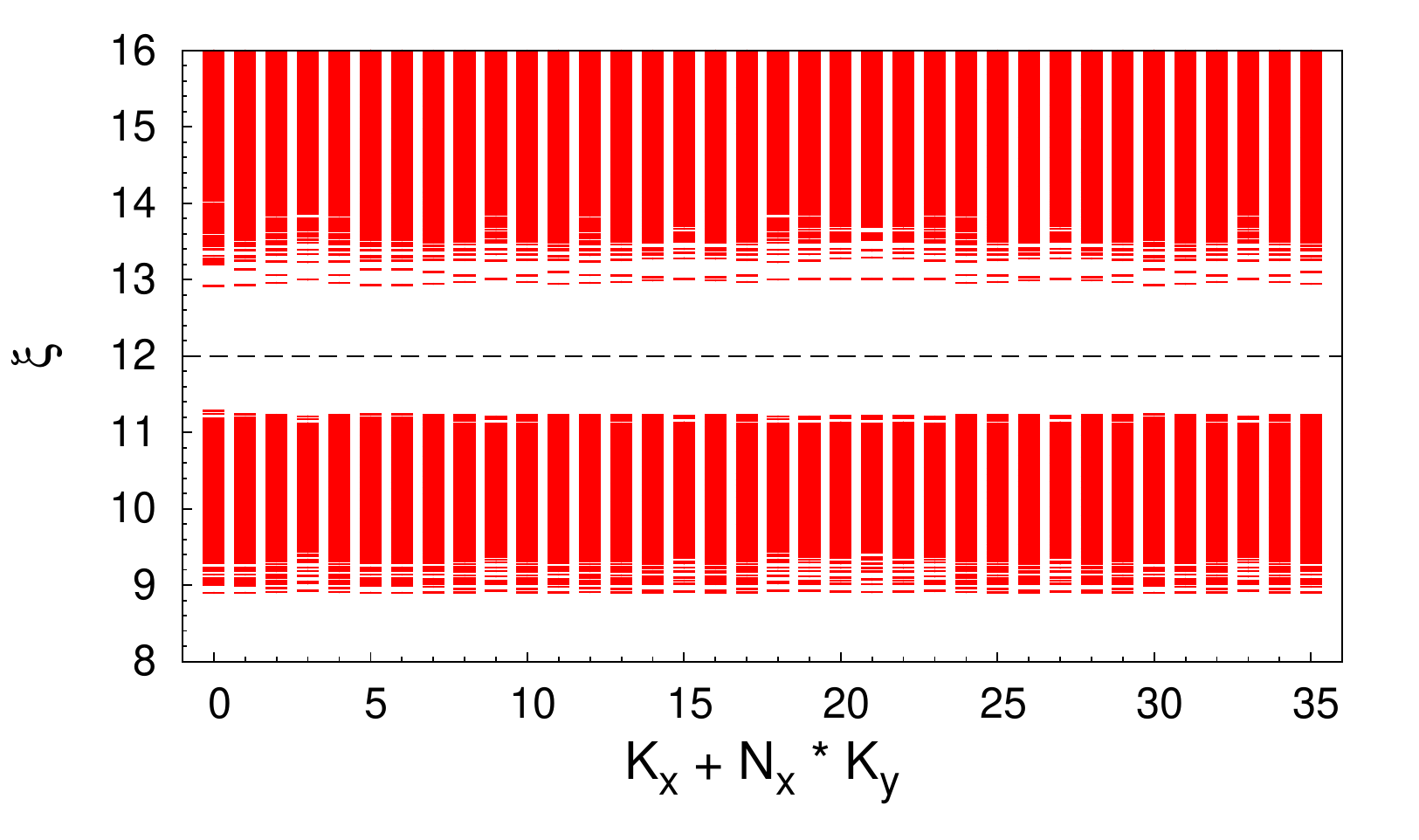}
\caption{Particle entanglement spectrum for $N=12, N_x=N_y=6$ and $N_A=4$. The $\xi$'s are the entanglement energies. The number of states below the dashed line is 741 in momentum sectors where $N_x \mod 2 = N_y \mod 2 = 0$ and 728 elsewhere. The total number below this line (26325) exactly matches the one predicted by the counting rule.}\label{entspec12}
\end{figure}

For the fractional Chern insulator, the situation is surprisingly much better: We observe a clear, large entanglement gap between low entanglement energy levels  and the high entanglement energy levels like those observed in Figs \ref{entspec10} and \ref{entspec12}. Moreover, the counting of the levels below the gap is identical to the counting of quasiholes of $N_A$ particles in $N_x \cdot N_y$ orbitals. In both examples shown in Figs. \ref{entspec10} and \ref{entspec12}, it matches per each momentum sector  that of $(1,3)$-admissible partitions of $N_A$ particles in $N_x \cdot N_y$ orbitals. We find it very revealing that the fractional Chern insulator ground states obtained here contain much clearer information (large, clear entanglement gap)  than the ground states of the Coulomb interaction in the FQH. The entanglement spectrum shows that the ground states by themselves contain essential information on the fractional nature of the excitations in the fractional Chern insulator. The current clean application of the entanglement spectrum also shows that this quantity is fundamentally useful toward revealing the physics of strongly-correlated states besides the usual FQH model wavefunctions for FQH states. 

The ES also provides some insight about the system when the number of sites in one direction is equal to 3. A clear entanglement gap is observed and the counting below this gap matches the expected one but only for $N_A=2$ and $N_A=3$. For larger values of $N_A$, the number of levels below the gap is lower than expected. This strongly suggest that for the small aspect ratios where the ration $N_y/N_x \rightarrow 0$, the state is not a fully developed fractional Chern insulator - or indeed, by our previous arguments, the single particle problem is not even a well-developed integer Chern insulator.  In the case of the other aspect ratios, all the ESs for any value of $N_A$ and any system sizes up to $N=12$ perfectly match the predicted counting. For the cases such as $N_x \times N_y = 6 \times 5$ and $6 \times 6$, the $(1,3)$ Pauli principle counting matches the ES data for \emph{each momentum sector} for all $N_A=1,2,3,4,5$. The perfect match clearly shows that this state is a Fractional Chern Insulator.

\section{Transition to the trivial insulator}

\begin{figure}[tbp]
\includegraphics[width=3.5in]{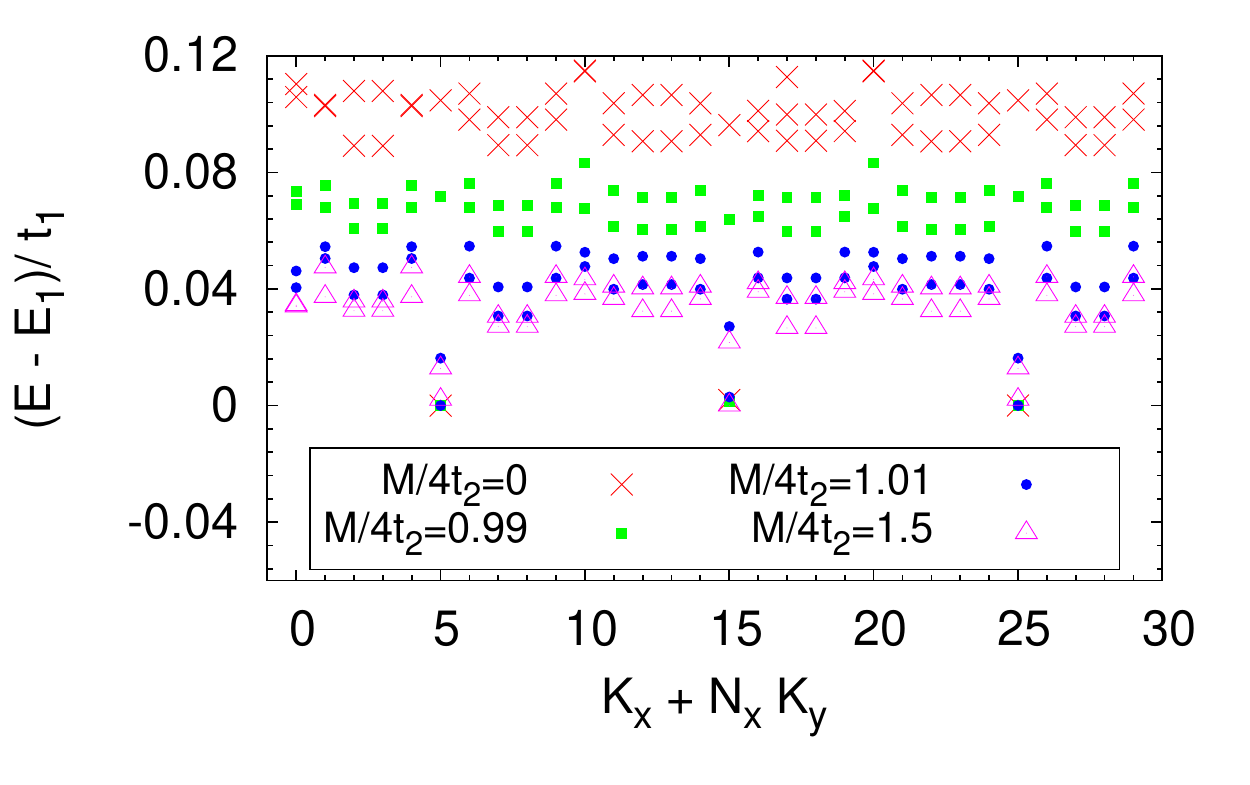}
\caption{Low energy spectrum for different values of $M$ at $N=10$, $N_x=5$ and $N_y=6$. We show only the two lowest energy per momentum sector. All energies are shifted such that the ground state energy is 0 for each $M$ value.}\label{spectrumtotrivial}
\end{figure}

\begin{figure}[tbp]
\includegraphics[width=3.5in]{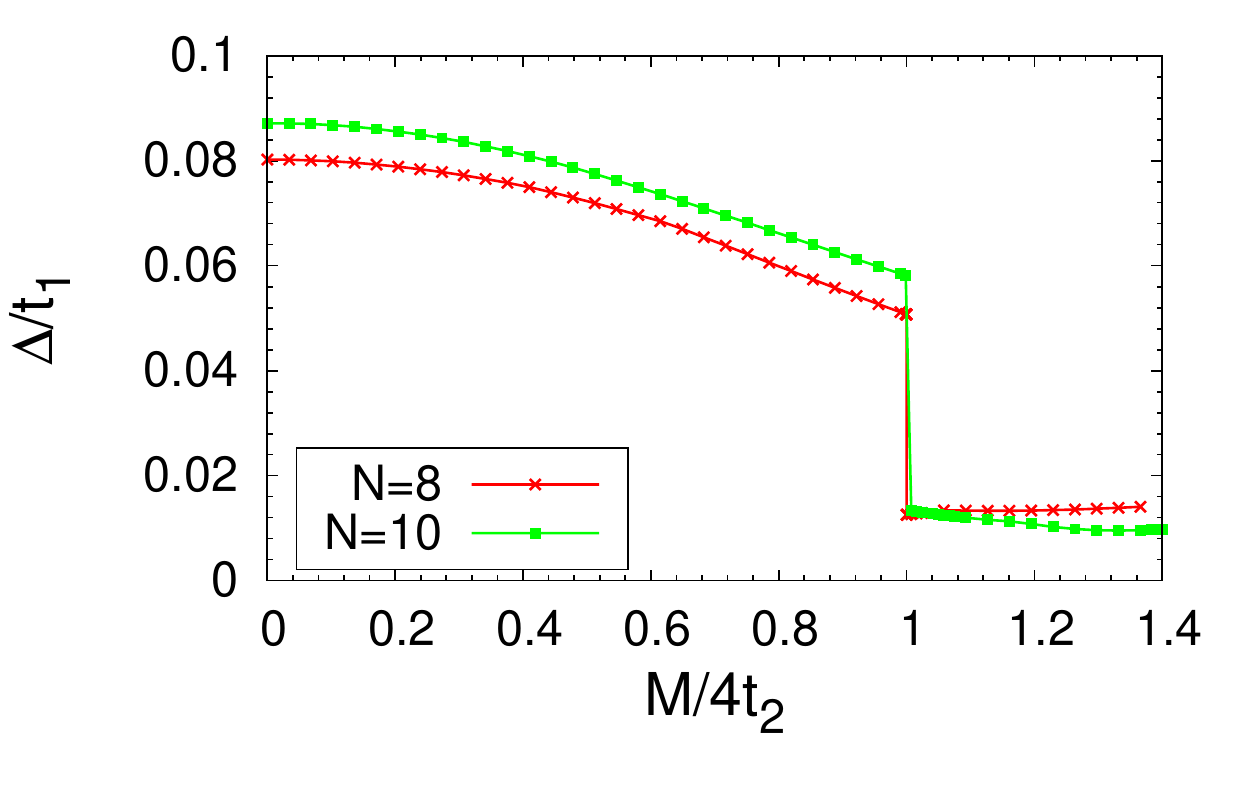}
\caption{Energy gap $\Delta$ as a function of $M$ at filling $\nu=1/3$ for $N=8$ and $N=10$ with $N_x=N/2, N_y=6$. For both system sizes, the transition occurs at $M = 4 t_2$.}\label{gaptotrivial}
\end{figure}

\begin{figure}[tbp]
\includegraphics[width=3.5in]{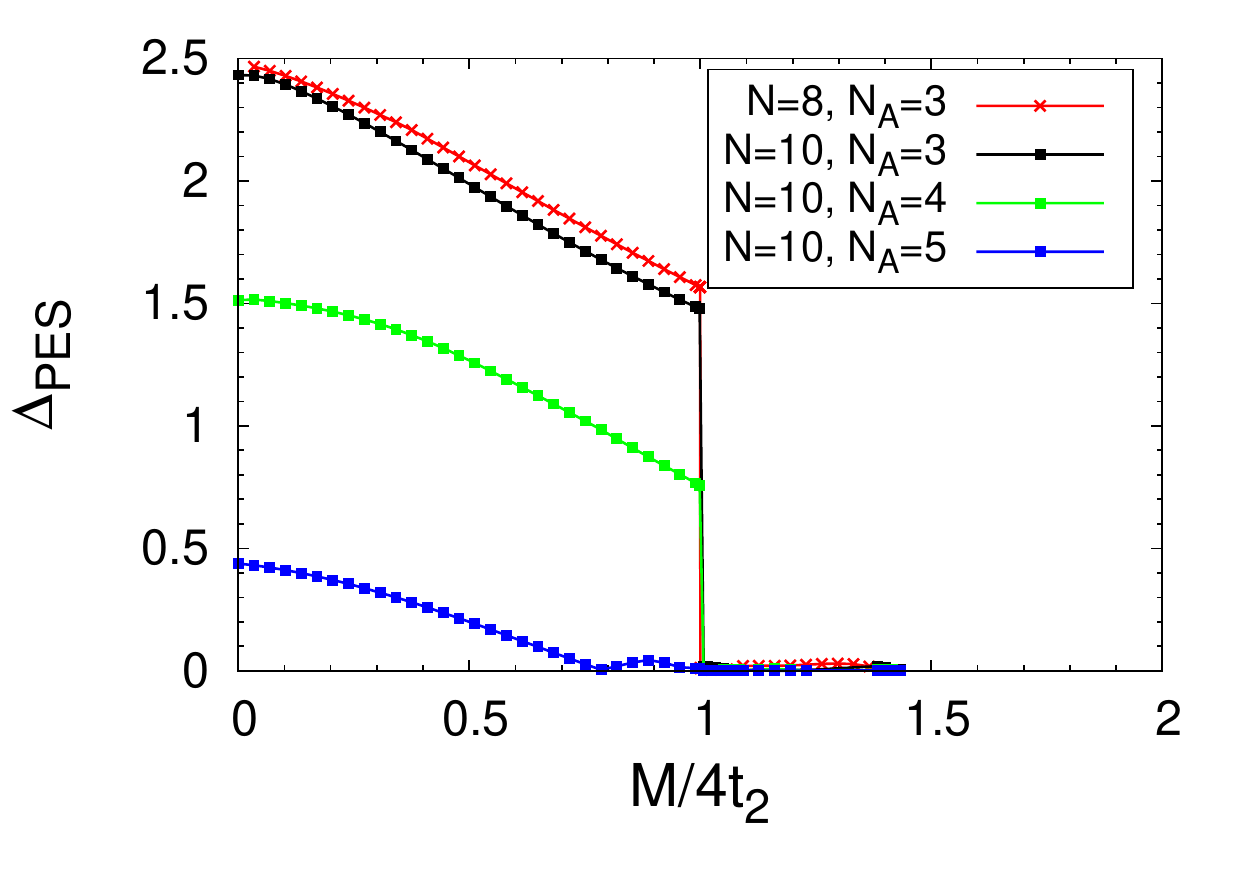}
\caption{Entanglement spectrum gap $\Delta_{\rm PES}$ as a function of $M$ at filling $\nu=1/3$ for $N=8$ and $N=10$ with $N_x=N/2, N_y=6$.}\label{pesgaptotrivial}
\end{figure}

Tuning the mass term above the $4t_2 $ threshold yields an insulator topologically equivalent to the atomic limit, with zero Hall conductance. Partially filling this insulator at say $M= 6 t_2$, we find no clear sign of a $3$-fold degenerate  ground state, and no clear gap to the excitation spectrum (see Fig.~\ref{spectrumtotrivial}).  In the interacting problem, we would think that the gap re-arrangement happens more quickly than $M=4 t_2$. This hypothesis is natural since the many-body gap is expected to be more sensitive than the single-particle one. The evolution of the many-body gap with the parameter $M$ is plotted in Fig.~\ref{gaptotrivial}. In the atomic limit $M \rightarrow \infty$, $B$ sites  have an energy $-M$ and all the particles are strongly localized on those sites. If all $B$sites  were occupied and $A$ sites unoccupied, the filling would be $1/2$, and hence as we are at a smaller filling $1/3$, and as the interaction couples sites $A$ and $B$, the ground state will be highly degenerate. We see that the many-body gap as a function of $M$ has a collapse at exactly the value of $M$ at which the one-body transition takes place.  We also plotted the entanglement gap as a function of $M$ in Fig.~\ref{pesgaptotrivial}. The entanglement gap is defined as the minimum separation over the different momentum sectors, between higher entanglement energy level belonging to the Laughlin counting and the lower entanglement energy level that does not belong to this counting. PES tracks the energy gap and its collapse also at the same value as the many-body and single-particle energy gaps. 

\section{Analytical Counting of Quasiholes Per Momentum Sector}

The heuristic generalized Pauli principle counting of the previous section gives the
correct (matching the data) total counting of quasiholes states manifold at any $N,
N_x, N_y$ (as long as the aspect ratio is two-dimensional). In most cases, although
not in all, it also gives the correct counting per each momentum sector. The
generalized Pauli principle is based on the correspondence between a $2$D momenta map to $1$D orbitals, which is heuristic. In this section we provide some analytic results of the counting
per momentum sector that has been tested to work for \emph{all} the momentum sectors
and is not based on a $2$-d to $1$-d folding. The counting of the total number
of quasihole states of a $1/3$ state of $N$ particles in $N_x \cdot N_y$ orbitals 
on the torus can be obtained by counting the number of   ways of putting $N$
particles in $N_x \cdot N_y$ orbitals with the restriction that there should be no more
than $2$ particles in $3$ consecutive orbitals ( $(1,3)$ admissible
partitions-generalized Pauli principle) on the sphere:
\beq
\#_{\text{qh sphere}}= \left(\begin{array}{cc}
N +  n\\
n \\
\end{array} \right),\;\;\;\;\; n= N_x N_y +2 - 3N
\eneq where $n$ is the number of quasiholes, and then subtracting the 
configurations that violate the $(1,3)$ generalized Pauli principle once we make the
system periodic to go to the torus. These configurations come in only three possible
ways and can be written in terms of the occupation number of orbitals 
\begin{eqnarray}
0100 (\text{all (1,3)-admissible configurations} ) 001 \nonumber \\
100 (\text{all (1,3)-admissible configurations} ) 0010 \nonumber \\
100 (\text{all (1,3)-admissible configurations} ) 001 
\end{eqnarray} These represent the only sets configurations that are allowed on the
sphere but that have more than one particle in $3$ consecutive orbitals if the
orbital space is made periodic. 
The number of the first two are equal by inversion symmetry and reads:
\beq
\left(\begin{array}{cc}
N -2+  n'\\
n' \\
\end{array} \right),\;\;\;\;\; n'= N_x N_y +1 - 3N
\eneq
while the third one reads
\beq
\left(\begin{array}{cc}
N -2+  n''\\
n'' \\
\end{array} \right),\;\;\;\;\; n''= N_x N_y +2 - 3N
\eneq The result then reads:
\beq
\#_{\text{qh torus}}=N_x N_y \frac{(N_x N_y - 2 N -1)!}{N!(N_x N_y - 3N)!}
\eneq This is the total number of quasiholes of a $1/3$ state on the torus
In a translationally invariant system, and for a FQH incompressible and featureless,
topological state, there is no particular reason why some momenta should have an
occupation number different from others. Indeed, both the interacting-particle and the
single-particle models are translationally invariant, and as such in the featureless
liquid ground state, momenta occupation numbers should be the same. Hence, on a
finite-size lattice for which we have $N_x \cdot N_y$ momenta, we would expect to
have $\frac{(N_x N_y - 2 N -1)!}{N!(N_x N_y - 3N)!}$ states for each momenta. While
this is always certainly true in the thermodynamic limit (as if the heuristic Pauli
principle), there are cases in finite size when  $\frac{(N_x N_y - 2 N -1)!}{N!(N_x
N_y - 3N)!}$ is not an integer. This means that in finite size cases, commensuration
effects weakly modify  the finite size counting of each sector so that the
requirement that there be an integer number of quasihole states for each momentum
sector is satisfied. This is the reason for the finite size changes of the number of
states in each momentum sector. 

Physical reasoning leads us further. When $N$ is relatively prime with both $N_x$
and $N_y$ there can be no commensuration effects, and the number of quasihole states
per each momentum sector has to be identical. Indeed, in this case, $\frac{(N_x N_y
- 2 N -1)!}{N!(N_x N_y - 3N)!}$ is an integer, is equal to the number of states per
momentum sector provided by our heuristic Pauli principle, and matches the numerical
data for the number of states per momentum sector. This is also the asymptotic value
of the number of states per momentum sector in the thermodynamic limit. 

Finite size effects become important in the case when the greatest common divisors
(GCD) of $N, N_x$ or $N, N_y$ is larger than $1$. Three cases can occur. First,
$GCD(N, N_y) >1, GCD(N, N_x)=1$: in this case, the number of quasihole states is the
same for all the $K_y$ total momentum that are divisible by $GCD(N, N_y)$ and is
different from the number of states at $K_y$ total momentum not divisible by $GCD(N,
N_y)$. A clear example of this situation is in Fig.~\ref{spectrumqh9}: $N=9,
N_x=5, N_y=6$ and hence $GDC(N,N_y)=3$. The momenta $(K_x,K_y)$ for which $K_y
\text{mod} 3=0$ have $19$ states in the quasihole subspace, whereas all momenta for
which $K_y \text{mod} 3 \ne 0$ have $18$ states in the quasihole subspace. This is
also equivalent to the counting of $(1,3)$ partitions obeying the generalized Pauli
principle.  The case $GCD(N, N_x) =1, GCD(N, N_x)>1$ is obviously the $x
\leftrightarrow y$ of the case presented above. Second when $GCD(N, N_y)  =GCD(N,
N_x)>1$, the number of quasihole states is the same for all $K_x, K_y$ which are
simultaneously \emph{both} divisible by $GCD(N, N_y) (=GCD(N, N_x) )$ and different
from the number of quasihole states at $K_x, K_y$ when either $K_x$ or $K_y$ is not
divisible by $GCD(N, N_y)$.  An example of this latter situation is in the counting
of the entanglement spectrum states of the $N=12$ ($N_x=6$, $N_y=6$) particles
ground state and $N_A=4$ in Fig.~\ref{entspec12}
The counting of states in the entanglement spectrum should be identical with that of
$N_A=4$ quasiholes in $N_x \cdot N_y = 36$ orbitals. As such, $GCD(N, N_x)= GCD(N, N_y)=2$.
Momenta $(K_x, K_y)=$ $(0,0)$, $(2,0)$, $(4,0)$, $(0,2)$, $(2,2)$, $(4,2)$, $(0,4)$,
$(2,4)$, $(4,4)$,   which are both divisible buy $GCD(N, N_x)$ have $741$ quasihole
states, whereas all other momenta, in which either $K_x$ or $K_y$ are not divisible
by $GCD(N, N_x)$ have $728$ quasihole states, for a total of $26325$ states, which
matches the total number of states of the $(1,3)$ Pauli principle. These rules for
the counting of states are physically motivated by the fact that momenta on the
lattice should be filled democratically except in the case when the number of
particles exhibits some commensuration (expressed by us as the condition $GCD(N,
N_{x,y})>1$) with the lattice dimensions. They have been checked for a range of particle number (up to $N=12$) and found to hold in all cases. The rules make sense
physically: In the case of commensuration, the momenta which are divisible by the
commensuration factor (GCD) should all exhibit the same counting (once the
commensuration condition has been established, the commensurate momenta are treated
democratically). This counting should be different from the counting of the momenta
incommensurate with the GCD. Whenever either $GCD(N, N_x)$ or $GCD(N, N_y)$ equal unity the Pauli principle counting of states matches the numerically found counting.

The third case, in which $GCD(N, N_y)  \ne GCD(N, N_x)\ne1$, has not been analyzed
and compared to the numerical data (as the size of the numerical computation becomes
too large). However, we can offer a conjecture for the quasihole state counting per
momentum. $(K_x, K_y)$ momenta which are not multiples of either $GCD(N, N_y)$ nor
$GCD(N, N_x)$ have identical counting, different from all others. The momenta for
which $GCD(N, N_x)$ divides $K_x$ but $GCD(N, N_y)$ does not divide $K_y$ all have
the same counting, different from any other. This is similar for   momenta for which $GCD(N,
N_y)$ divides $K_y$ but $GCD(N, N_x)$ does not divide $K_x$. Finally, momenta for which
$K_x$ is divisible by $GCD(N, N_x)$ and $K_y$ is divisible by $GCD(N, N_y)$  have
same counting, different from the rest.

\section{Conclusion}

In conclusion, we have shown that the ground state of the $\nu=1/3$ flat-band Chern insulator in the presence of repulsive interactions is an incompressible state with Hall conductance $1/3$, and that quasihole excitations satisfying fractional statistics. The presence of the excitation whose counting is identical to those in the FQH $\nu=1/3$ state is a clear and up-to-now missing proof that the ground state is indeed a fractional Chern insulator and not a $1/3$ charge-density wave state. We have presented a mapping between the lattice momenta and the torus orbitals and shown that the counting and total momenta of the ground states and those of the quasiholes can be obtained by employing a generalized Pauli principle of not more than $2$ particles in $3$ consecutive orbitals. We have then shown that the entanglement spectrum of the ground state also has an entanglement gap and that the levels below the gap match in counting those of the Laughlin quasihole states.

\section{Acknowledgements} 
BAB wishes to thank T.L. Hughes, F.D.M. Haldane, S. Sondhi  and C. Chamon and 
L. Santos for very useful discussions.
BAB was supported by Princeton Startup Funds, Sloan Foundation, NSF DMR-095242, and NSF China 11050110420, and a MRSEC grant at Princeton University, NSF DMR-
0819860

\section{Appendix I: Quasiholes and the Generalized Pauli Principle}

The counting of quasiholes per momentum sector is slightly more complicated, and we have not found the full counting rule per momentum sector. Although the total number of quasiholes always matches the counting of partitions, the map between the two-dimensional momentum numbers and the one-dimensional orbital partition that allowed us to find the momentum of the ground states cannot always work for the quasihole states. This can be easily seen as the partition map between the momentum of the $i'th$ particle $k_{xi}, k_{yi}$ and the one-dimensional momentum $(1,3)$-admissible partition $\lambda =(\lambda_1,\ldots, \lambda_N)$ with  $\lambda_i= N_x \cdot k_{yi }+k_{xi}$ does not (for quasiholes) have to respect the inversion symmetry that the model studied has. To illustrate this, take two particles $i,j$, their momenta $(k_{xi}, k_{yi})$, $(k_{xj}, k_{yj})$ and their orbital momentum $\lambda_i, \lambda_{j}$ \emph{satisfying the $(1,3)$ admissibility} $|\lambda_i - \lambda_j|\ge 3$. The inversion property acts at the single-particle level and transforms $k_x, k_y \rightarrow -k_x,-k_y$. To make the inverted momentum belong to the positive numbers, we must add $N_x, N_y$ to the $x,y$ momenta except if the momenta are $0$, in which case we do nothing. With these new momenta, we form the inverted partitions $\lambda_{i}^{inv},\lambda_{j}^{inv}$ and check whether these violate the admissibility rule, i.e. whether $|\lambda_{i}^{inv} - \lambda_{j}^{inv}| <3$. We then have to analyze seven cases separately out of which three cases turn out to be problematic: 
\begin{enumerate}
\item  $k_{xi}\ne0, k_{yi}\ne 0, k_{xj}\ne0, k_{yj}\ne 0 $, $\lambda_i^{inv} = N_x N_y + N_x - \lambda_i$,  $\lambda_j^{inv} = N_x N_y + N_x - \lambda_j$ and  $|\lambda_{i}^{inv} - \lambda_{j}^{inv}|  = |\lambda_{i} - \lambda_{j}| >3 $  - inversion symmetry preserved. 
\item  $k_{xi}=0, k_{yi}\ne 0, k_{xj}\ne0, k_{yj}\ne 0 $, $\lambda_i^{inv} = N_x N_y - \lambda_i$,  $\lambda_j^{inv} = N_x N_y + N_x - \lambda_j$ and  $|\lambda_{i}^{inv} - \lambda_{j}^{inv}|  = |N_x+\lambda_{i} - \lambda_{j}|  $ - and hence inversion symmetry is not preserved in the cases, $k_{yj}= k_{yi}, k_{xj}=N_x-2, N_x-1$ or $k_{yj}= k_{yi}+1, k_{xj}=1,2$; 
\item  $k_{xi}=0, k_{yi}\ne 0, k_{xj}\ne0, k_{yj}=0 $, $\lambda_i^{inv} = N_x N_y  - \lambda_i$,  $\lambda_j^{inv} = N_x - \lambda_j$ and  $|\lambda_{i}^{inv} - \lambda_{j}^{inv}|  = |N_x N_y - N_x -\lambda_{i} + \lambda_{j}|$ - and hence inversion symmetry is not preserved in the cases $k_{yi}=N_y-1, k_{xj}=1,2$; 
\item  $k_{xi}=0, k_{yi}\ne 0, k_{xj}= 0, k_{yj}\ne 0 $, $\lambda_i^{inv} = N_x N_y - \lambda_i$,  $\lambda_j^{inv} = N_x N_y - \lambda_j$ and  $|\lambda_{i}^{inv} - \lambda_{j}^{inv}|  = |\lambda_{i} - \lambda_{j}| >3 $  - inversion symmetry preserved. 
\item  $k_{xi}=0, k_{yi}= 0, k_{xj}\ne0, k_{yj}\ne 0 $, $\lambda_i^{inv} = \lambda_i=0$,  $\lambda_j^{inv} = N_x N_y + N_x - \lambda_j$ and  $|\lambda_{i}^{inv} - \lambda_{j}^{inv}|  = |N_y N_x + N_x +\lambda_{i} - \lambda_{j}| >3 $ if $N_x>1$  - inversion symmetry preserved. 
\item  $k_{xi}=0, k_{yi}=0, k_{xj}\ne0, k_{yj}= 0 $, $\lambda_i^{inv} =\lambda_i=0$,  $\lambda_j^{inv} = N_x - \lambda_j$ and  $|\lambda_{i}^{inv} - \lambda_{j}^{inv}|  <3 $  - for $\lambda_j = k_{xj} =N_x-2, N_x-1$ then inversion symmetry is not preserved.  
\item  $k_{xi}=0, k_{yi}= 0, k_{xj}=0, k_{yj}\ne 0 $, $\lambda_i^{inv} = \lambda_i=0$,  $\lambda_j^{inv} = N_x N_y  - \lambda_j$ and  $|\lambda_{i}^{inv} - \lambda_{j}^{inv}|  = |N_x N_y - \lambda_{j}| >3 $  - inversion symmetry preserved.
\end{enumerate}

Despite the above, in most (but not all) of the cases, the counting of states per momentum is \emph{still} given by the $(1,3)$ -partitions of the same total momentum $K_x = \sum_{i=1}^N k_{xi}$  and  $K_y = \sum_{i=1}^N k_{yi}$, even though there is no inversion symmetry in the partitions (as proven above; we stress that the total counting of quasiholes always matches that of $(1,3)$ partitions, as it should for a Laughlin state; only the $2D$ to $1$D mapping proposed in this paper sometimes fails for quasihole states). What happens in these cases is that momentum sectors related by inversion symmetry both contain the same number of partitions which under the inversion operation are not $(1,3)$ admissible. So the effect of the $2$D to $1$D mapping on the sectors not respecting the inversion symmetry is magically canceled. We find that the matching per momentum sector of the exact diagonalization states and the $1.3$ - admissible partitions to be a rather remarkable demonstration of the $\nu=1/3$ character of the states. 

\bibliography{FractionalChernInsulator.bib}

\end{document}